\documentclass[ba,linksfromyear,nameyear]{Custom}

\usepackage{bm}
\usepackage[normalem]{ulem}
\usepackage{mathtools}
\usepackage{booktabs}
\usepackage{multirow}
\usepackage{natbib}
\usepackage{amsfonts}
\usepackage{url}

\begin{document}
\begin{frontmatter}
\title{The Attraction Indian Buffet Distribution}
\runtitle{The Attraction Indian Buffet Distribution}

\begin{aug}
\author[addr1]{\fnms{Richard L.} \snm{Warr}\corref{}\thanksref{t1}\ead[label=e1]{warr@byu.edu}},
\author[addr2]{\fnms{David B.} \snm{Dahl}},
\author[addr3]{\fnms{Jeremy M.} \snm{Meyer}},
\and
\author[addr4]{\fnms{Arthur} \snm{Lui}}

\runauthor{R. Warr et al.}

\address[addr1]{Department of Statistics, Brigham Young University, 2152 WVB, Provo, UT 84602}
\address[addr2]{Department of Statistics, Brigham Young University, 2152 WVB, Provo, UT 84602}
\address[addr3]{Department of Statistics, Brigham Young University, 2152 WVB, Provo, UT 84602}
\address[addr4]{Statistical Sciences Group, Los Alamos National Laboratory, Los Alamos, NM 87545}

\thankstext{t1}{Corresponding author.}
\end{aug}

\begin{abstract}
We propose the attraction Indian buffet distribution (AIBD), a distribution
for binary feature matrices influenced by pairwise similarity information.
Binary feature matrices are used in Bayesian models to uncover latent variables
(i.e., features) that explain observed data. The Indian buffet process
(IBP) is a popular exchangeable prior distribution for latent feature matrices.
In the presence of additional information, however, the exchangeability
assumption is not reasonable or desirable. The AIBD can incorporate pairwise
similarity information, yet it preserves many properties of the IBP, including
the distribution of the total number of features. Thus, much of the interpretation
and intuition that one has for the IBP directly carries over to the AIBD.
A temperature parameter controls the degree to which the similarity information
affects feature-sharing between observations. Unlike other nonexchangeable
distributions for feature allocations, the probability mass function of
the AIBD has a tractable normalizing constant, making posterior inference
on hyperparameters straight-forward using standard MCMC methods. A novel
posterior sampling algorithm is proposed for the IBP and the AIBD. We demonstrate
the feasibility of the AIBD as a prior distribution in feature allocation
models and compare the performance of competing methods in simulations
and an application.
\end{abstract}


\end{frontmatter}

\vspace*{-3.5pt}\section{Introduction}\vspace*{-3.5pt}
\label{intro}

Two primary functions for data modeling are to relate observed data to
each other and to future observations. These purposes of modeling imply
that the data (both previously observed and yet to be observed) are, to
some extent, related to each other. Thus, when modeling, we assume that
the observed data are somehow interconnected and possess some information
about future observations. These relationships are often complex and not
easily captured in traditional models. Bayesian nonparametric latent feature
models account for these complexities by allowing any number of features
to connect observations to one another, without assuming a predetermined
relationship structure.\looseness=-1

One prior for Bayesian nonparametric latent feature models is the Indian
buffet process (IBP) \citep{griffiths2011indian}. In a realization
of the IBP, an observation may possess zero, one, or any number of features
possibly shared with the other observations. The total number of possible
features is unbounded, and can\vadjust{\goodbreak} theoretically account for any amount of
complexity in the data. Under the Bayesian construct, the IBP is used as
a prior distribution for a feature allocation, and is updated with data
(via a likelihood) which results in a posterior distribution of that feature
allocation.

A major assumption of the IBP is that all observations are exchangeable.
In other words, before the data are collected one item is indistinguishable
from another. This assumption can be quite restrictive if,
\textit{a priori}, information about the observations is known. For example,
the amount of trade between pairs of countries might be known, yet this
information cannot be incorporated into a model which insists on an exchangeable
feature allocation distribution.

To account for distance information between observations,
\cite{gershman2014distance} developed the distance dependent Indian buffet
process (dd-IBP). This method allows a modeler to indicate,
\textit{a priori}, the distances between each pair of observations. In this
paper we also propose a generalization of the IBP which incorporates pairwise
distances into the feature allocation prior, namely the attraction Indian
buffet distribution (AIBD). However, the AIBD retains a few desirable characteristics
of the IBP which are lost with the dd-IBP. The first is that our method
retains the same number of expected features as the IBP, whereas the dd-IBP
changes the number of expected features, with respect to the IBP. The AIBD
also has a tractable probability mass function (pmf) which readily allows
for standard MCMC techniques on hyperparameters. Another property of the
AIBD is that the expected number of shared features between two customers
can increase or decease (in relation to the IBP). In the dd-IBP the expected
number of shared features typically decreases as the distances are included.
The methods associated with the AIBD are implemented in the package
\textit{aibd} available on the comprehensive R archive network (CRAN). We feel, and demonstrate in detail, that
the characteristics of our proposed method provide specific advantages
over the dd-IBP.

The organization of the paper is as follows. In Section~\ref{sec:litrev} we discuss some of the previous work for this methodology
and establish the notation and models needed for this article. In Section~\ref{sec:aibd} we present the AIBD pmf. Next, in Section~\ref{sec:aibdprior}, we investigate key properties of the AIBD and compare
them to the dd-IBP. Then, in Section~\ref{post_samp}, we outline a new
recipe for posterior simulation when using an IBP or AIBD prior. Section~\ref{dataAnalysis} is a description of a classification analysis of an
Alzheimer's disease neuroimaging study \citep{dinov2009efficient},
and demonstrates the advantages of using an AIBD prior. We finish in Section~\ref{discussion} with a brief summary of this work.\vspace*{-3.5pt}

\section{Literature Review}\vspace*{-3.5pt}
\label{sec:litrev}

In this section we discuss the primary literature needed for our proposed
method and define notation used in this article. A short discussion of
the Chinese restaurant process is included as an aide for those who might
be familiar with random partition models, but are new to feature allocation
models.\vspace*{-3.5pt}

\subsection{The Chinese Restaurant Process}\vspace*{-3.5pt}

Bayesian nonparametric models seek to capture latent structure in data.
In clustering applications where each observation is assigned to a group\vadjust{\goodbreak}
to form a partition, the Chinese restaurant process (CRP) serves as a prior
distribution over all possible partitions. The CRP resembles a Chinese
restaurant with an infinite number of tables, in which $n$ customers enter
one at a time. Each customer picks a table to sit at, favoring tables with
more customers. The resulting assignment of customers to tables induces
a partition of the customers. Thus, the CRP will create latent features
that are exclusive. The CRP is exchangeable. Consequently, the probability
of any two customers being in the same cluster is the same for all pairs
of customers.

However, the constraint that each customer has an equal chance of being
clustered with any other customer may not fully reflect existing (\textit{a
priori}) knowledge. Certain covariates like socioeconomic background, age,
or other distances in time and space, will likely impact the clustering.
Therefore, instead of constraining all datapoints to be equidistant at
the start of the analysis, it could be useful to have an expert incorporate
pairwise distances into the prior. \cite{ddCRP} developed the distance-dependent
Chinese restaurant process (ddCRP) to facilitate these distances
\textit{a priori}. However, the ddCRP does not have a tractable probability
mass function (pmf), which makes using standard Markov chain Monte Carlo
(MCMC) techniques for posterior inference on hyperparameters difficult.

\cite{EwensDist} proposed the Ewens-Pitman attraction distribution (EPA),
which also allows pairwise distance information to be included in the CRP.
This distribution has an explicit pmf with a tractable normalizing constant
and pmf. This is ideal for using standard MCMC sampling methods. Like the
dd-CRP, the EPA places more probability on partitions that group similar
items. But unlike the ddCRP, the EPA does not change the distribution of
the number of subsets; it only influences how the datapoints are clustered
together within the class of partitions having the same number of subsets.\looseness=-1

Similar to how the EPA incorporated distance information while preserving
many of the qualities of the CRP, we propose a new distribution, the AIBD,
that incorporates pairwise distance information in the IBP prior. Although
the existing dd-IBP uses pairwise distances, we propose a distribution
that preserves some properties and intuition for the IBP and has an explicit
pmf with a tractable normalizing constant.\vspace*{-3.5pt}

\subsection{The Indian Buffet Process}\vspace*{-3.5pt}

A popular prior distribution for Bayesian nonparametric latent feature
allocation models is the Indian buffet process
\citep{ghahramani2006infinite}. The Indian buffet process (IBP) puts a
prior distribution on feature allocations. The generative construct of
the IBP can be thought of as an Indian buffet restaurant with a seemingly
infinite number of dishes. A fixed number of customers enter the buffet
one at a time to sample dishes. The first customer enters and takes a Poisson($
\alpha $) number of unique dishes. After the first customer, the
$i^{th}$ customer samples each existing dish with probability
$m_{k}/i$, where $m_{k}$ is the number of customers who have previously
sampled dish $k$. The $i^{th}$ customer then takes Poisson($\alpha /i$)
new dishes. Thus, popular dishes will tend to be taken more often by later
customers and the number of new dishes to be sampled will diminish as more
customers enter the restaurant.

The dishes taken by each customer can be encoded in a (binary) feature
allocation matrix $\bm{Z}$ where rows and columns\vadjust{\goodbreak} correspond to customers
and dishes, respectively. In this matrix, $z_{i,k} = 1$ indicates that
customer $i$ took dish $k$. Likewise $z_{i,k} = 0$ indicates that customer
$i$ did not take dish $k$. $\bm{Z}$ also describes how the customers
share features. For example, $z_{i,k}=z_{j,k}=1$ indicates customers
$i$ and $j$ both took (i.e. share) dish $k$. The dishes are analogous to
latent features and thus customers who share more dishes are thought to
share similar (unobserved) attributes. Although, technically, an infinite
number of dishes are not sampled (which are represented as columns of zeros),
these are generally removed from $\bm{Z}$.

Since the dishes are indistinguishable, the ordering of the columns in
$\bm{Z}$ is irrelevant. As a result, any permutation of the columns
in $\bm{Z}$ will correspond to the same feature allocation. Considering
all column-permutations of $\bm{Z}$ that represent the same feature allocation
is important. One way to map each $\bm{Z}$ to its unique feature allocation
is to consider the equivalence class of matrices in left-ordered form.
A left-ordered form ($\mathit{lof}$) matrix can be obtained by taking the
binary number of each column (with the most significant digit in the first
row) and then ordering the columns in descending order from left to right.
A $\bm{Z}$ in left-ordered form will thus have a stair-like pattern,
with the first 1 appearing in a column only when a new dish is taken. To
take into account the indistinguishable columns, we add a combinatoric
term to the probability mass functions in (\ref{ibp-pmf}) and
(\ref{aibd:pmf}). Thus a specific $\bm{Z}$ will refer to the class of all
feature allocations that map to the same left-ordered form.

The expected number of sampled dishes per customer is the mass parameter
$\alpha $, a positive real number. We will denote $N$ as the number of
customers and $K$ as the total number of dishes taken by at least one customer.
That is, the matrix $\bm{Z}$ has $N$ rows and $K$ nonzero columns. Define
$x_{i}$ as the number of new dishes customer $i$ takes, $y_{i}$ as the
number of sampled dishes before customer $i$, and $m_{-i,k}$ as the number
of customers that took dish $k$ before customer $i$. For convenience, let
$H_{N} = \sum _{i}^{N}(1/i)$ be the $N^{th}$ harmonic number. The IBP pmf
is shown in (\ref{ibp-pmf}) and can be loosely divided into 3
pieces: the combinatorial term, the Poisson term, and the Bernoulli (binary)
term. The cardinality of all possible non-zero binary columns of length
N is $2^{N}-1$, so $\prod _{h=1}^{2^{N}-1} K_{h} !$ iterates over the sample
space of all distinct non-zero columns in $\bm{Z}$. Where $K_{h}$ represents
the number of columns for the $h^{\text{th}}$ possible configuration. By
following the constructive pattern, the IBP pmf can be expressed as
\begin{align}
\begin{split}
\label{ibp-pmf}
\large \text{P}(\bm{Z}|\alpha )=& \Bigg [
\frac{\prod _{i=1}^{N} x_{i} !}{ \prod _{h=1}^{2^{N}-1} K_{h} !}
\Bigg ]
\frac{\alpha ^{K}\exp \{-\alpha H_{N}\}}{\prod _{i=1}^{N}(i^{x_{i}}~x_{i}!)}\times\\
&\prod \limits _{i=2}^{N}\prod \limits _{k=1}^{y_{i}}\left (
\hspace{.39em}
\frac{m_{-i,k}}{i}
\hspace{.39em}
\right )^{z_{i,k}} \left (
\hspace{.30em}
1-\frac{m_{-i,k}}{i}
\hspace{.30em}
\right )^{1-z_{i,k}}.
\end{split}
\end{align}
Note that the $\prod _{i=1}^{n} x_{i}!$ term will cancel, but it is not
removed so one can intuitively see the origin of the various parts of the
pmf. In the case where no dishes have been sampled before customer
$i$ enters ($y_{i} = 0$), the result of the double product is defined to
be 1.\looseness=-1

The IBP prior has the property that customers are exchangeable, i.e. changing
the order of the rows in $\bm{Z}$ has no impact on the probability
of any given feature allocation. As a result, the expected number of shared
features for all customers is uniform. Therefore, on average, customers
will share the same number of features. While this may be desirable in
instances where nothing is known about the customers, additional
\textit{a priori} information may be relevant and should be used to influence
how features are shared. For example, in the context of the Indian buffet
restaurant, we may know that certain customers have similar dietary preferences
before they walk in the restaurant. In other applications, time-based,
spatial, or covariate dependencies can be used to create prior dependencies
between data points. Instead of assuming customers are exchangeable before
the analysis, it may be helpful to relax this assumption in light of additional
information.\looseness=-1

\subsubsection{The Distance Dependent IBP}
\label{ddIBPIntro}

\cite{gershman2014distance} proposed a generalization of the IBP, the distance
dependent Indian buffet process (dd-IBP), which incorporates pairwise distance
information, e.g. distance in time, in space, or computed from covariates.
Under the dd-IBP, customers that are ``nearer'' share dishes more frequently,
whereas customers that are further apart share dishes less frequently.
This behavior is achieved as follows. First, customers either ``own'' a
dish, or are connected to customers (including themselves) that own a dish.
The number of dishes that customer $i$ ``owns'' is
$\lambda _{i} \sim \text{Poisson}(\alpha /h_{i})$, where $\alpha $ is again
a mass parameter, $N$ is the number of customers,
$h_{i} = \sum _{j=1}^{N} f(d_{i,j})$, $d_{i,j}$ denotes the distance between
customers $i$ and $j$, and $f(d)$ is a monotone decay function satisfying
$f(0)=1$ and $f(\infty )=0$. The set of dishes ``owned'' by customer
$i$ are labelled (arbitrarily) as
$\mathcal{K}_{i} = (\sum _{j<i}\lambda _{j}, \sum _{j\le i} \lambda _{i}]$,
and the total number of owned dishes is
$K=\sum _{i=1}^{N} \lambda _{i}$. Customer $i$ then ``connects'' to customer
$j$ for dish $k \in \{1,\dots ,K\}$ with probability
$a_{i,j} = f(d_{i,j}) / h_{i}$. By connecting, customer $i$ ``inherits''
dishes owned or inherited by customer $j$. A dish is automatically inherited
by the dish owner. Customers that inherit the same dishes thus share features.
An $N\times K$ connectivity matrix $\bm{C}$ encodes customer connections,
in which $c_{i,k}=j$ denotes that customer $i$ connects to customer
$j$ for dish $k$. Based on $\bm{C}$, a feature allocation
$\bm{Z}$ is deterministically computed. Note that the relationship
between $\bm{C}$ and $\bm{Z}$ is many-to-one.

The joint distribution for the connectivity matrix $\bm{C}$ and
$K$-dimensional dish ownership vector $\bm{c}^{\star }$, where
$c_{k}^{\star }\in \{1,\dots ,N\}$, denotes the customer that owns dish
$k$ is given by\looseness=-1
\begin{eqnarray}
\Pr (\bm{C}, \bm{c}^{\star }\mid \bm{D}, \alpha , f) &=&
\Pr (\bm{c}^{\star }\mid \alpha ) \cdot \Pr (\bm{C}\mid
\bm{c}^{\star }, \bm{D}, f), ~\text{with}
\\
\Pr (\bm{c}^{\star }\mid \alpha ) &=& \prod _{i=1}^{N} \Pr (
\lambda _{i} \mid \alpha ), ~\text{and}
\label{eq:owner-prob}
\\
\Pr (\bm{C} \mid \bm{c}^{\star }, \bm{D}, f) &=& \prod _{i=1}^{N}
\prod _{k=1}^{K} a_{i, c_{i,k}},
\end{eqnarray}\looseness=0
where $\bm{D}$ is a $N\times N$ distance matrix. Note that in~\eqref{eq:owner-prob},
$\bm{c}^{\star }$ is deterministic given the $\lambda $'s. Finally, the
probability of a feature allocation matrix $\bm{Z}$ is
\begin{eqnarray}
\Pr (\bm{Z}\mid \bm{D}, \alpha , f) = \sum _{(\bm{c}^{\star }, \bm{C}):\phi (\bm{c}^{\star }, \bm{C})=\bm{Z}}
\Pr (\bm{c}^{\star }, \bm{C} \mid \bm{D}, \alpha , f),
\label{eq:ddibp-pmf-Z}
\end{eqnarray}
where $\phi (\bm{c}^{\star }, \bm{C})$ maps a given connectivity
matrix and ownership vector to a feature allocation matrix
$\bm{Z}$. Crucially, the pmf of the feature allocation
$\bm{Z}$ in~(\ref{eq:ddibp-pmf-Z}) requires marginalizing over
all possible connectivity matrices and ownership vectors. This is intractable
when $N$ is moderately large. Thus, in practice, posterior sampling algorithms
for making inference on $\bm{Z}$ rely on sampling from the posterior
distribution of $\bm{C}$ and $\bm{c}^{\star }$.

The dd-IBP reduces to the IBP when the proximity matrix is a lower diagonal
matrix of 1's. When this is not the case, neither the distribution of the
number of dishes per customer nor the distribution of the total number
of features $K$ are the same as that of the IBP. Since the AIBD will also
incorporate pairwise distance information, we will compare properties of
both the dd-IBP and AIBD in Section~\ref{sec:aibdprior}.

\subsubsection{Recent Applications of and Other Work on the IBP}

The value of the IBP can be seen in its repeated application in research
problems, particularly, in the biological sciences. See, for example,
\cite{hai2011inferring,chen2013phylogenetic,xu2013nonparametric,sengupta2014bayclone,xu2015mad,lee2015bayesian,lee2016bayesian,ni2018bayesian, lui2020bayesian}.

In terms of methodological extensions, other work has been done to relax
the exchangeability constraint of the IBP in the literature.
\cite{dIBP} proposed the dependent IBP, which introduces dependence through
a hierarchical Gaussian process. \cite{miller} proposed a generalization
of the IBP, the phylogenetic Indian buffet process, that introduces dependencies
between objects by conditioning on a dependency tree. This also reduces
down to the IBP when all branches meet at the root. This method performs
well for data with genealogical relationships and expresses prior object
similarity through a tree. The Indian buffet Hawkes process (\citealp{tan})
extended the IBP to capture latent temporal dynamics by incorporating ideas
from the Hawkes process. \cite{williamson2020new} presents a class of nonexchangeable
dynamic models constructed by adapting the IBP. These models are tailored
to data that are believed to be generated by latent features exhibiting
temporal persistence. We focus our comparison on the dd-IBP since it, like
our AIBD, introduces dependence through pairwise distances.

\subsection{The Linear Gaussian Latent Feature Model (LGLFM)}
\label{sec_lglfm}

The typical likelihood in the Bayesian nonparametric literature for latent
feature models is the linear Gaussian latent feature model (LGLFM). Using
similar notation as found in \cite{griffiths2011indian}, the LGLFM is defined
as:
\begin{equation}
\bm{X} = \bm{Z}\bm{A} + \bm{\varepsilon },
\end{equation}
where $\bm{X}$ is an $N \times D$ matrix of $N$ observations on $D$ variables.
$\bm{Z}$ is an $N \times K$ binary matrix of $0$s and $1$s and indicates
which features are turned off or on for a specific observation (i.e., row
of $\bm{X}$). $\bm{A}$ is a $K \times D$ matrix whose rows are the latent
features and whose prior is a matrix Gaussian distribution, with probability
density function
\begin{equation}
p(\bm{A}|\sigma _{A}) \propto \exp \left \{  -
\frac{1}{2\sigma ^{2}_{A}} \text{trace}\left (\bm{A}^{T}\bm{A} \right )
\right \}  .
\end{equation}
Finally, $\bm{\varepsilon }$ is an $N \times D$ matrix and represents the
error term of the model; it also has a matrix Gaussian distribution similar
to $\bm{A}$ but has different dimensions and its parameter is
$\sigma _{X}$. Technically $\bm{Z}$ and $\bm{A}$ have an infinite number
of columns and rows (respectively). However, only $K$ columns of
$\bm{Z}$ are non-zero. Thus the zero columns of $\bm{Z}$ are discarded
along with the associated rows of $\bm{A}$ and we treat those matrices
as if they have a finite number of rows and columns (see
\cite{ghahramani2005infinite} or \cite{griffiths2011indian} for more details).
If $\bm{A}$ is integrated out of the model, the collapsed likelihood is:
\begin{align}
\begin{split}
p(\bm{X}|\bm{Z},\sigma _{X},\sigma _{A}) \propto {}&
\frac{1}{\sigma _{X}^{ND-KD}\sigma _{A}^{KD}\left |\bm{Z}^{T}\bm{Z}+\frac{\sigma _{X}^{2}}{\sigma _{A}^{2}}\bm{I}\right |^{D/2}}
\times
\\
& \exp \left \{  -\frac{1}{2 \sigma _{X}^{2}} \text{trace} \left (
\bm{X}^{T} \left [ \bm{I} -\bm{Z} \left ( \bm{Z}^{T}\bm{Z}+
\frac{\sigma _{X}^{2}}{\sigma _{A}^{2}} \bm{I} \right )^{-1} \bm{Z}^{T}
\right ] \bm{X} \right ) \right \}  .
\end{split}
\label{colapLik}
\end{align}
We use this collapsed likelihood in the posterior inference section with
various feature allocation priors on $\bm{Z}$ in Section~\ref{post_samp}.

\section{The Attraction Indian Buffet Distribution}
\label{sec:aibd}

We propose a generalization of the IBP, the attraction Indian buffet distribution
(AIBD). We describe how we obtain this distribution by modifying the generative
model of the IBP to include distance information between customers. Incorporating
existing distance information about the customers, in turn, influences
how the dishes are shared. We then show the nonexchangeable probability
mass function and compare it to the IBP.

Distance information can be stored in a symmetric $N \times N$ pairwise
distances matrix. The distance $d_{i,j}$ between customers $i$ and
$j$ is located in the $i^{\text{th}}$ row and $j^{\text{th}}$ column. We
transform these distances to similarity values, where 0 indicates negligible
similarity and larger values indicate a greater similarity between customers.
Various transformations can appropriately map distance to similarity and
a temperature parameter $\tau $ is introduced to accentuate the effect
of these distances. In general, we require: i. the transformation function
$f(\tau , d_{i,j})$ to be a monotonically decreasing function in
$d_{i,j}$ for fixed $\tau $, ii.
$f(\tau _{1},d_{1})f(\tau _{2},d_{2}) \leq f(\tau _{2},d_{1})f(\tau _{1},d_{2})$
for $d_{1} \leq d_{2}$ and $\tau _{1} \leq \tau _{2}$, and iii. when
$\tau =0$, it must return a constant in the interval $(0,\infty )$. These
properties imply that, as the temperature $\tau $ increases, the ratio
$f(\tau , d_{1}) / f(\tau , d_{2})$ increases for fixed
$d_{1} \leq d_{2}$. Thus, increasing $\tau $ accentuates the effect of
distance on similarity. Valid decay functions include those listed in
\cite{gershman2014distance}. For example, the constant function,
$f(\tau , d) = c$, where $c$ is a positive constant; the exponential function,
$f(\tau , d)=\exp (-\tau d)$; the reciprocal function,
$f(\tau , d) = (d + \nu )^{-\tau }$, with shift $\nu > 0$ added to avoid
division by 0; and the window function,
$f(\tau , d) = \mathbf{1}(d \le 1/\tau )$. The result of the element-wise
transformations of the distance matrix is a similarity matrix. The AIBD
uses the similarity matrix $\boldsymbol{\Lambda }$ to incorporate dependence
between customers. A temperature parameter is also used in the dd-IBP and
a few functions to transform distances to proximities are suggested in
\cite{gershman2014distance}. It is worth mentioning that, unlike the AIBD,
the dd-IBP does not require a symmetric distance matrix.

Due to non-exchangeability, the AIBD is conditioned on a permutation parameter
$\bm{\rho }$, which is any permutation vector of the integers 1 to
$N$. This controls the order in which customers arrive and allows us to
characterize temporal or spatial dependence \textit{a priori}. In many cases,
however, the data has no natural ordering. That is, it may not make sense
to say the data depends on the order it was observed or recorded. For this
reason, the permutation parameter is typically averaged out of the model
by Monte Carlo integration or enumeration. A reasonable prior distribution
for $\bm{\rho }$, is the uniform distribution over all possible permutations
of the integers 1 to $N$.

Because we desire to preserve many of the properties of the IBP, the AIBD
has a generative model very similar to the IBP. Using the same restaurant
analogy as the IBP, the AIBD can also be thought of as an Indian buffet
restaurant where customers enter one at a time. Like the IBP, the first
customer takes a Poisson($\alpha $) number of dishes and the
$i^{th}$ customer takes Poisson($\alpha / i$) new dishes. However, instead
of sampling existing dishes with probability proportional to the number
of customers who have already sampled the dish, the AIBD also uses pairwise
similarity information. The $i^{th}$ customer gets existing dishes with
probability equal to the sum of similarities of individuals who have that
dish, divided by the sum of the total similarity with all previous individuals,
all multiplied by $(i-1)/i$. When all the pairwise similarities are the
same, the probability of sampling existing dishes reduces to that in the
IBP. Thus, the IBP can be thought of as a special case of the AIBD when
all the pairwise similarity components are identical.

By following the constructive process described above, the pmf of the AIBD
can be obtained, as shown in (\ref{aibd:pmf}). The AIBD is a distribution
over feature allocations, therefore, the support is over all binary matrices
with $N$ rows and only non-zero columns. Since the ordering of features
does not matter, this probability mass function returns the probability
of all feature allocations that are equivalent to the supplied
$\bm{Z}$. The AIBD uses a pairwise distance matrix $\bm{D}$ and is conditioned
on the permutation vector $\bm{\rho }$ of the integers 1 to $N$. The parameters
and notation carry the same meaning as they do in the IBP pmf in~(\ref{ibp-pmf}).
The pmf of the AIBD is
\begin{align}
\begin{split}
\label{aibd:pmf}
\text{P}(\bm{Z}|\alpha , \bm{\rho }, \tau ) ={}&
\frac{\prod _{i=1}^{N} x_{i} !}{ \prod _{h=1}^{2^{N}-1} K_{h} !}
\cdot
\frac{\alpha ^{K} \exp \{-\alpha H_{N}\}}{\prod _{i=1}^{N}(i^{x_{i}}~x_{i}!)}\times\\
&\prod \limits _{i=2}^{N}\prod \limits _{k=1}^{y_{i}} \left (
\frac{h_{ik}(\tau ) \cdot (i-1)}{i} \right )^{z_{i,k}} \left ( 1-
\frac{h_{ik}(\tau ) \cdot (i-1)}{i} \right )^{1-z_{i,k}},
\end{split}
\end{align}
where $h_{ik}(\tau )$ is defined as
\begin{equation}
\label{hik}
h_{ik} =
\frac{\sum \limits _{j=1}^{i-1}f(\tau , d_{\rho _{j},\rho _{i}}) \cdot z_{j,k}}{\sum \limits _{j=1}^{i-1}f(\tau , d_{\rho _{j},\rho _{i}})}.
\end{equation}
The term $d_{\rho _{j},\rho _{i}}$ corresponds to the distance between
the $i^{th}$ and $j^{th}$ individuals in a given permutation
$\bm{\rho }$.

The IBP and AIBD priors have the same support, and the probability mass
functions are fairly similar. The key differences are the probabilities
defined in the double product of \eqref{ibp-pmf} and \eqref{aibd:pmf}. The IBP has a customer sample a dish proportional to
the number of times it has been taken. Customers in the AIBD also sample
popular dishes more frequently, but the probability is also dependent on
similarity information. Note that when $f(\tau , d)$ is constant,
$h_{i,k} = m_{-i, k} / (i - 1)$, and thus, the terms in the double product
of \eqref{aibd:pmf} reduce to $(m_{-i,k} / i)$ and
$(1 - m_{-i, k} / i)$ as in the IBP p.m.f. in \eqref{ibp-pmf}. When $f$ is not constant, $h_{i,k}$ is larger (or smaller)
when a greater (or fewer) number of ``similar'' customers have previously
sampled dish $k$.

\section{Properties of the AIBD}
\label{sec:aibdprior}

In this section, we explore some of the properties of the AIBD and compare
them to the IBP and dd-IBP. A distribution on possible $\bm{Z}$'s implies
a distribution on the number of non-zero columns. The distribution of the
number of non-zero columns in the AIBD is the same as that in the IBP because,
in the constructive model, the distance information is not used to determine
the number of dishes. Thus, the distribution of features is invariant to
similarity information included in the AIBD. We will compare this result
by simulation to the dd-IBP, where the distribution of the number of features
changes with the temperature parameter and the distance information. We
will also compare how the features are shared between customers as a function
of temperature for both the AIBD and dd-IBP. Thus, we will proceed by focusing
on the total number of features and number of shared features in
$\bm{Z}$.

\subsection{Distribution of the Number of Features}
\label{sec:numFeat}

In the IBP and AIBD, the distribution of the number of features $T$ (i.e.
number of non-zero columns in the $\bm{Z}$ matrix) can be explicitly characterized.
Since a new column in $\bm{Z}$ is generated when a customer samples a
new dish, the total of the number of features is equal to the sum of the
number of new dishes each customer takes. From the generative model of
the IBP and AIBD, let the number of new dishes that the $i^{th}$ customer
takes be $X_{i}= $ Poisson($\alpha /i$). Since the Poisson draws are independent
between customers, the total number of features,
$T = \sum _{i}{X_{i}}$ is distributed
\begin{equation}
T \stackrel{\mathclap{\normalfont \tiny \mbox{d}}}{=} T_{IBP}
\stackrel{\mathclap{\normalfont \tiny \mbox{d}}}{=} T_{AIBD} \sim
\text{Poisson}(\alpha H_{N}),
\label{nf_ibd}
\end{equation}
where $H_{N}$ is the $N^{th}$ harmonic number. This distribution is identical
for the IBP and AIBD because the similarity information present in the
AIBD is only used to determine how existing features are shared. No new
dishes or columns are generated based on the distance information. Note
that this is also invariant to the permutation parameter $\bm{\rho }$. The
distribution of the number of features can only be changed by adjusting
the mass parameter $\alpha $ or changing the number of customers $N$.

The generative model for the dd-IBP, however, is different in that the
proximity, like the AIBD's similarity information, changes the total number
of features. The dd-IBP uses a proximity matrix $\bm{P}$ to capture
\textit{a priori} pairwise distance information. Using the dd-IBP generative
model in \cite{gershman2014distance},\vadjust{\goodbreak} the number of new features for customer
$i$ is $X_{i} \sim \text{Poisson}(\alpha /h_{i})$. Thus,
\begin{equation}
\label{nf_ddibp}
T_{dd\text{-}IBP} = \sum \limits _{i=1}^{N}X_{i} \sim \text{Poisson}
\left (\alpha \sum _{i=1}^{N}\frac{1}{h_{i}}\right ),
\end{equation}
where $h_{i} = \sum _{j=1}^{N} P_{i,j}$ and $P_{i,j}$ corresponds
to the proximity measure between customers $i$ and $j$. Thus,
$h_{i}$ is the sum of the $i^{th}$ row in the dd-IBP proximity matrix.
The proximity matrix in the dd-IBP differs slightly from the similarity
matrix $\bm{\Lambda }$ in the AIBD. The dd-IBP requires self-proximity to
be 1 and infinite distances to be mapped to a proximity of 0. Thus, only
monotonic transformations that map distance values from
$[0,\infty )$ to $[0,1]$ can be used. We will employ the same transformation
mentioned earlier, $f(\tau , d_{i,j})=\exp (-\tau d_{i,j})$, where
$d_{i,j}$ corresponds to the pairwise distance matrix used for the AIBD.
The proximity matrix is called \textit{sequential} if when $i < j$,
$P_{i,j}=0$; that is, when the proximity matrix is lower-diagonal.
This allows current customers to only inherit dishes from previous customers.
For the dd-IBP to simplify to the IBP, one condition is that
$\bm{P}$ must be sequential, analogous to how the IBP restaurant analogy
only allows customers to enter one at a time.

\begin{table}
\begin{minipage}[t]{.49\textwidth}
\begin{center}
\textbf{\small {AIBD Similarity Matrix ($\tau =1$)}}
\begin{tabular}{rrrrr}
\hline 1 & 2 & 3 & 4 & 5 \\\hline
1.00 & 0.89 & 0.51 & 0.02 & 0.02 \\
0.89 & 1.00 & 0.55 & 0.03 & 0.02 \\
0.51 & 0.55 & 1.00 & 0.04 & 0.03 \\
0.02 & 0.03 & 0.04 & 1.00 & 0.36 \\
0.02 & 0.02 & 0.03 & 0.36 & 1.00 \\
\end{tabular}
\end{center}
\end{minipage}
\begin{minipage}[t]{.49\textwidth}
\begin{center}
\textbf{\small {dd-IBP Proximity Matrix ($\tau =1$)}}
\begin{tabular}{rrrrr}
\hline 1 & 2 & 3 & 4 & 5 \\ \hline
1.00 & 0.00 & 0.00 & 0.00 & 0.00 \\
0.89 & 1.00 & 0.00 & 0.00 & 0.00 \\
0.51 & 0.55 & 1.00 & 0.00 & 0.00 \\
0.02 & 0.03 & 0.04 & 1.00 & 0.00 \\
0.02 & 0.02 & 0.03 & 0.36 & 1.00 \\
\end{tabular}
\end{center}
\end{minipage}
\caption{Similarity matrix $\bm{\Lambda }$ and sequential proximity matrix
$\bm{P}$ at a fixed temperature $\tau $ using the natural permutation. Although
both have ones on the diagonal, it is only required for the dd-IBP's proximity
matrix $\bm{P}$.}
\label{sim_matrices}
\end{table}

The dd-IBP simplifies to the IBP when the proximity matrix is a lower-diagonal
matrix of 1's. When this happens, $h_{i}$ in (\ref{nf_ddibp})
is equal to $i$, so $T_{dd\text{-}IBP}$ and $T_{IBP}$ have the same distribution.
A comparison of an AIBD similarity matrix using the natural permutation
(integers 1 through N in ascending order) and a sequential dd-IBP proximity
matrix is shown in Table~\ref{sim_matrices}. Both matrices were generated
from the USArrests dataset in R. We selected the states New Hampshire,
Iowa, Wisconsin, California, and Nevada respectively. We then calculated
the pairwise Euclidean distances of the 5 states after centering and scaling
the covariates. For the AIBD, we put the pairwise distances in a
$5\times 5$ matrix $\bm{D}$ as described in Section~\ref{sec:aibd}. Finally,
to get the similarity or proximity, we applied the transformation exp($-
\tau d_{i,j}$) to each distance element. Note that in this example, the
individual states are analogous to customers in the Indian buffet restaurant.\looseness=-1

\begin{figure}
\includegraphics[scale=.4]{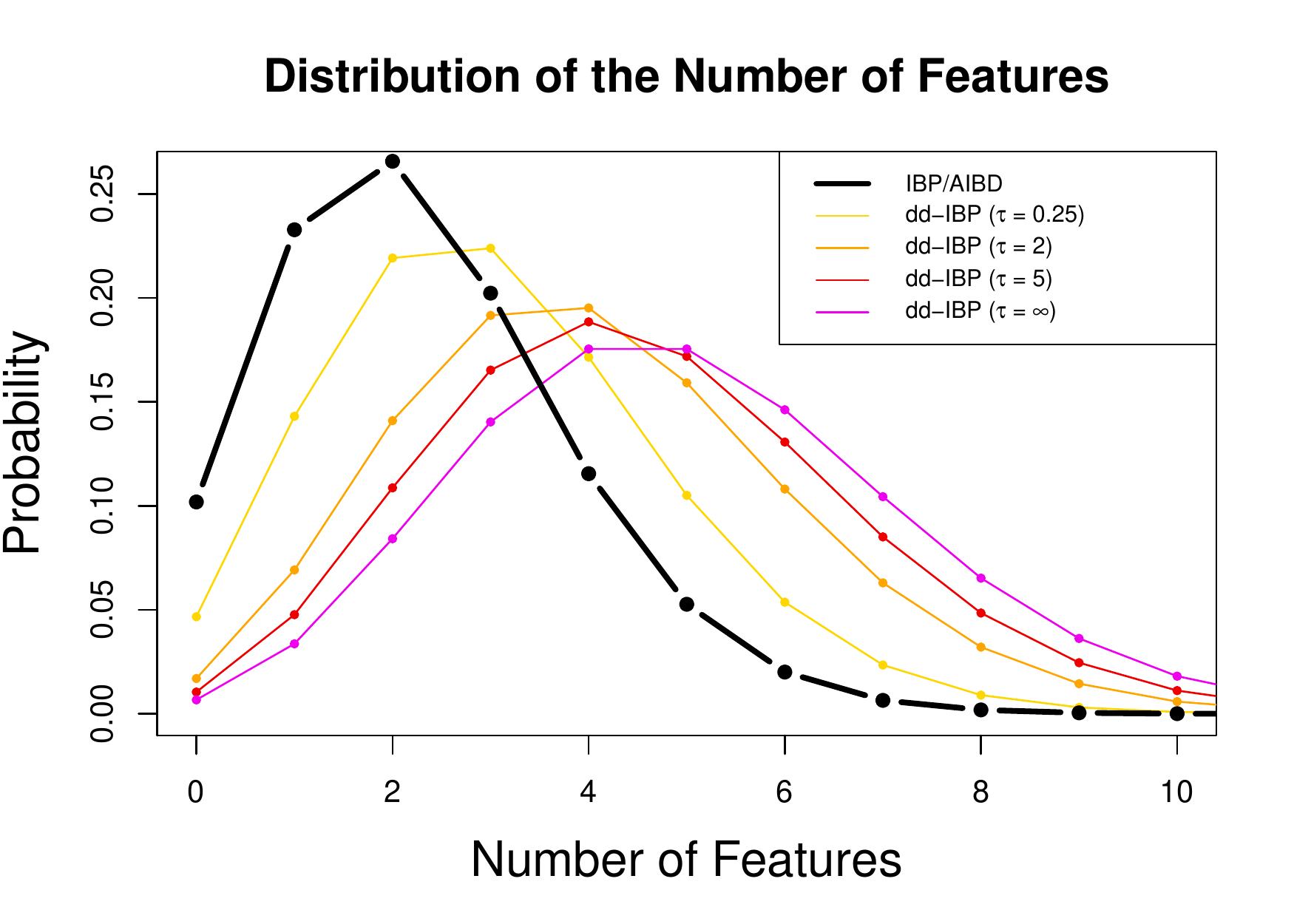}
\caption{The distribution of $T_{AIBD}$ and $T_{IBP}$ is displayed as the bold
black line. The distribution of $T_{dd\text{-}IBP}$ is displayed in the narrow
colored lines for various temperatures. This figure shows that the distribution
of the number of features for the dd-IBP (using the proximity matrix in
Table~\ref{sim_matrices} and when $\tau >0$) is stochastically greater than the
number of features for the IBP and the AIBD.}
\label{nfeat}
\end{figure}

After fixing $\alpha =1$, $N=5$, and using the permutation of increasing
natural numbers 1 through $N$, we obtain the probability distribution of
the number of features as shown in Figure~\ref{nfeat}. Recall that both
$T_{AIBD}$ and $T_{IBP}$ have the same distribution as $T$, whereas the
number of non-zero columns in the dd-IBP, $T_{dd\text{-}IBP}$, varies by
temperature and distance information. The distributions of $T$ and
$T_{dd\text{-}IBP}$ were given in (\ref{nf_ibd}) and (\ref{nf_ddibp}).
Figure~\ref{nfeat} illustrates that, for this proximity matrix, the dd-IBP
has a higher number of expected features than the IBP. As the temperature
increases, $T_{dd\text{-}IBP}$ has a limiting distribution of a Poisson($
\alpha N$), and values on the off diagonal of the dd-IBP's proximity matrix
approach zero. Thus, the proximity matrix $\bm{P}$ approaches an identity
matrix, and $\alpha \sum _{i}^{n}(1/h_{i}) \rightarrow \alpha N$ as
$\tau \rightarrow \infty $. In this case, the $N$ customers do not share
features and individually sample $\alpha $ dishes, on average. On the other
hand, the AIBD preserves the same distribution of the number of features
as the IBP, regardless of temperature, and is only affected by the mass
parameter. This is an important attribute of the AIBD, because as the number
of features increases so does the computational complexity of inference.
The AIBD also encourages more feature sharing than the dd-IBP, as discussed
in the next section.

Another property the AIBD preserves from the IBP is that they have the
same expected number of features per customer, which is $\alpha $ (see
Section~\ref{sec:aibd-exp-features-per-row} in the supplementary material
for a proof). These two properties ensure that the average number of features
and the average number of features per customer are identical between the
AIBD and the IBP. Therefore, the IBP and the AIBD do the same amount of
feature sharing, which is solely controlled by the parameter
$\alpha $. However, the AIBD differs from the IBP, not by changing the
amount of feature sharing, but by allowing the propensity of customers
to share a feature to depend on their pairwise distances.

\subsection{Expected Number of Shared Features}
\label{aibd_enumf}

One consequence of exchangeability in the IBP is that the expected number
of shared features is identical for all customer pairs. This is not a desirable
property when, \textit{a priori}, one knows that a pair of customers are
more alike when compared to another customer. The AIBD is able to include
this information; which has the desired effect of changing the expected
number of shared features for a pair of customers, while the average overall
feature sharing remains the same as with the IBP.

\begin{figure}[]
    \begin{minipage}[]{.65\textwidth}
        \includegraphics[width=3.65in]{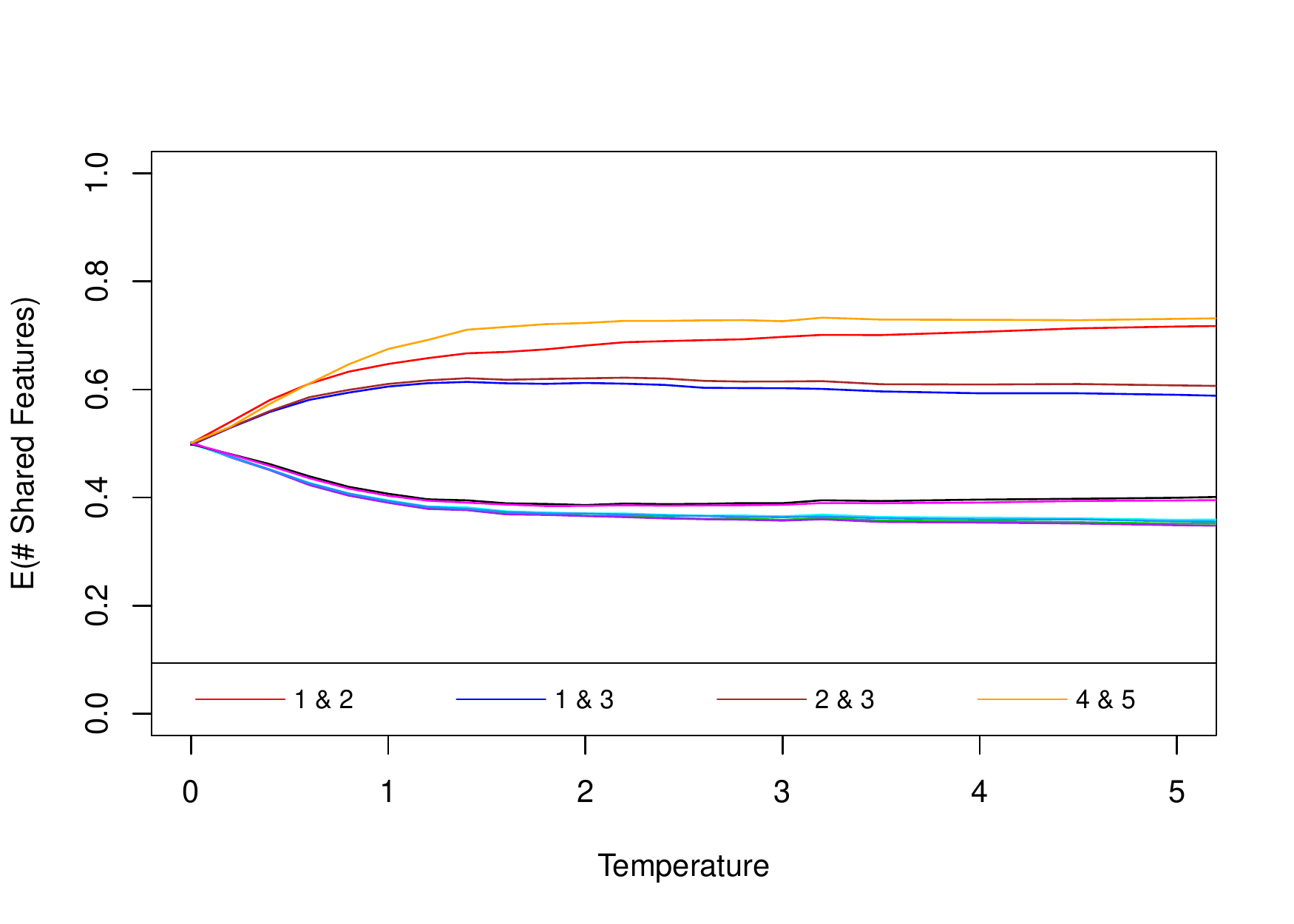}
    \end{minipage}
    \begin{minipage}[b]{.02\textwidth}
    \end{minipage}
    \begin{minipage}[b]{.33\textwidth}
\begin{center}
\textbf{\small {AIBD Similarity Matrix ($\tau =1$)}}
\begin{tabular}{rrrrr}
\hline 1 & 2 & 3 & 4 & 5 \\\hline
1.00 & 0.89 & 0.51 & 0.02 & 0.02 \\
0.89 & 1.00 & 0.55 & 0.03 & 0.02 \\
0.51 & 0.55 & 1.00 & 0.04 & 0.03 \\
0.02 & 0.03 & 0.04 & 1.00 & 0.36 \\
0.02 & 0.02 & 0.03 & 0.36 & 1.00 \\
\end{tabular}
\end{center}
	\end{minipage}
     \caption{The expected number of shared features as a function of temperature in the AIBD (averaged over all permutations) with corresponding similarity matrix when $\tau=1$. Each line represents a customer pair, with some pairs identified in the legend. This figure shows that the AIBD adjusts feature sharing for pairs of customers given the similarity matrix; the IBP would have all customers share on average 0.5 features.}
    \label{temp_plot}
\end{figure}

In the AIBD, customers that are closer in distance tend to share more features.
The degree to which customers share features can be adjusted by the temperature
parameter. An example is shown in the plot of Figure~\ref{temp_plot}. The
plot was obtained by fixing $\alpha =1$ and by using pairwise similarity
information found in Table~\ref{sim_matrices}. We fixed $\alpha $ because
changing the mass parameter only scales the y-axis in the plot of Figure~\ref{temp_plot}.
The lines indicate the expected number of shared features for a pair of
customers.

When the temperature is zero, the AIBD reduces to the IBP, and so all customer
pairs have the same expected number of shared features. Due to variability
between permutations, each line in Figure~\ref{temp_plot} was calculated
by averaging the expected number of shared features across all possible
$N!=120$ permutations. In other words, $\bm{\rho }$ was integrated out after
placing a uniform prior on it. Since $N$ is small, we enumerated across
all permutations up to 7 possible features, which accounted for 99.4\%
of the probability mass.

A similarity matrix from a fixed temperature is shown to the right of the
feature sharing plot in Figure~\ref{temp_plot}. As expected, customers
with higher similarity tend to share more features. However, even though
customers 1 and 2 have the highest similarity, customers 4 and 5 tend to
share more features. This behavior occurs by design, the expected number
of features per customer in the AIBD is the same as in the original IBP,
so the distances do \emph{not} effect the \emph{number} of features, just
\emph{how} features are shared among each other. So, the issue is not which
$\exp \{-\tau d_{i,j}\}$ is the largest or smallest, but rather the relative
sizes of these quantities. See (\ref{aibd:pmf}) and (\ref{hik}).
As an illustration, consider the similarity matrix in Figure~\ref{temp_plot} and suppose that customer 5 is the last to be allocated
features. Among customers 1--4, customer 5 will highly favor sharing features
with customer 4 since the relative similarity
$0.36 / ( 0.02 + 0.02 + 0.03 + 0.36 ) = 0.84$ is close to 1. Conversely,
if customer 1 is the last to be allocated, its relative attraction to customer
2 is only $0.89 / ( 0.89 + 0.51 + 0.02 + 0.02 ) = 0.61$, which is not as
large as 0.84. Therefore, it would be expected that even though customers
1 and 2 have the highest similarity (i.e., 0.89), customers 4 and 5 tend
to share more features. The sharing arrangements will depend somewhat on
the permutation since, \textit{a priori}, feature allocation only depends
on the similarities of previously allocated customers, but the overall
effect remains that number of shared features among pairs of customers
will be driven by relative rather than absolute similarities. Table~\ref{Tab:ExpectVsSim} in the supplementary material compares the expected
number of features between two states with the similarity values for this
example with three different temperatures.

We now examine the effect of increasing $N$ on the expected number of shared
features per customer in the AIBD. We do this using plots similar to Figure~\ref{temp_plot}.
Additionally, we use plots for both the AIBD and dd-IBP to compare how
they each influence feature sharing. We use sample sizes $N=5$ and
$N=50$ to demonstrate the effect of increasing $N$. Since it is computationally
infeasible to enumerate $50!$ permutations, we create the plots using Monte
Carlo estimation.

\begin{figure}[t!]
\includegraphics[width=\textwidth]{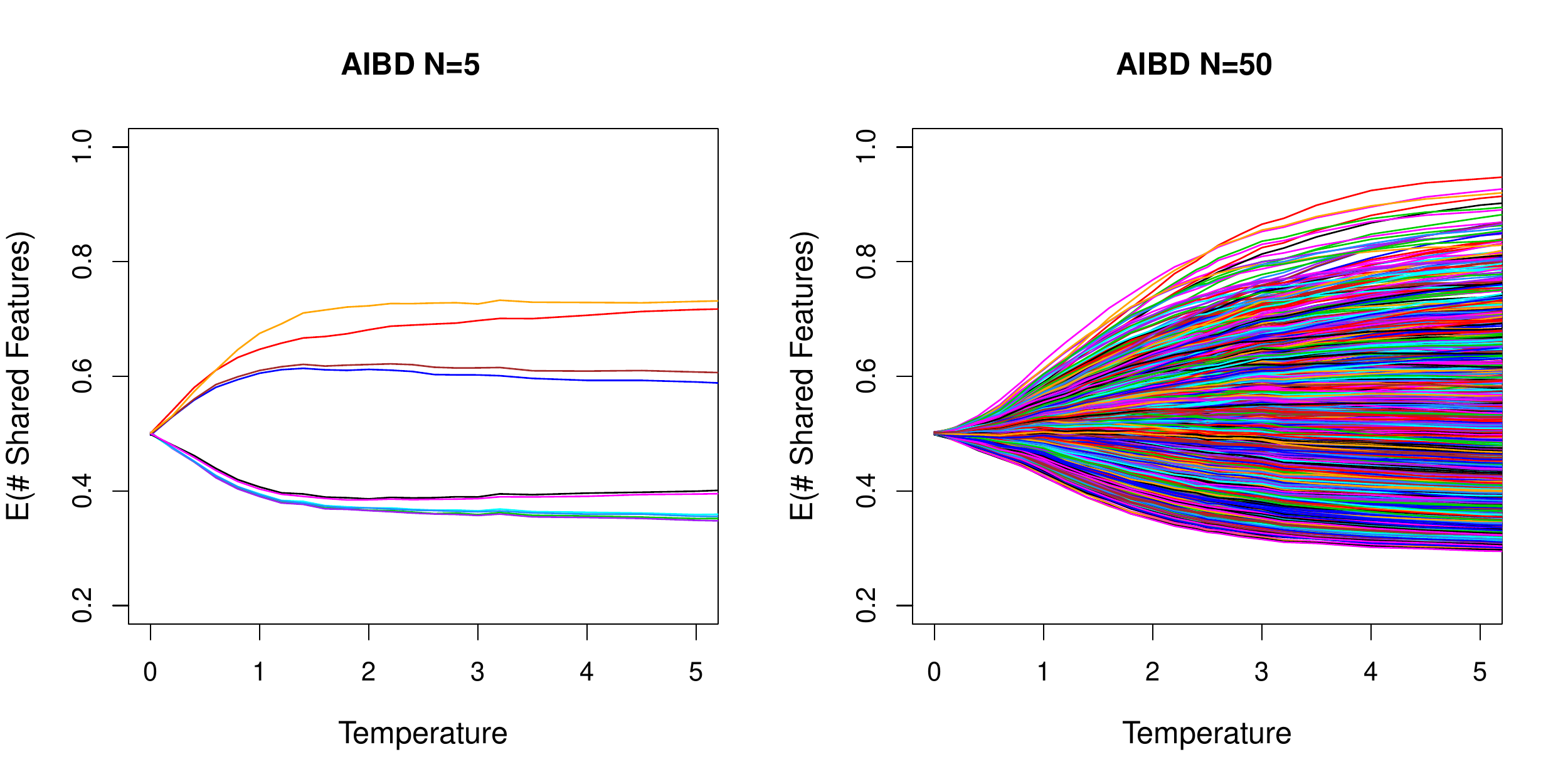}
\includegraphics[width=\textwidth]{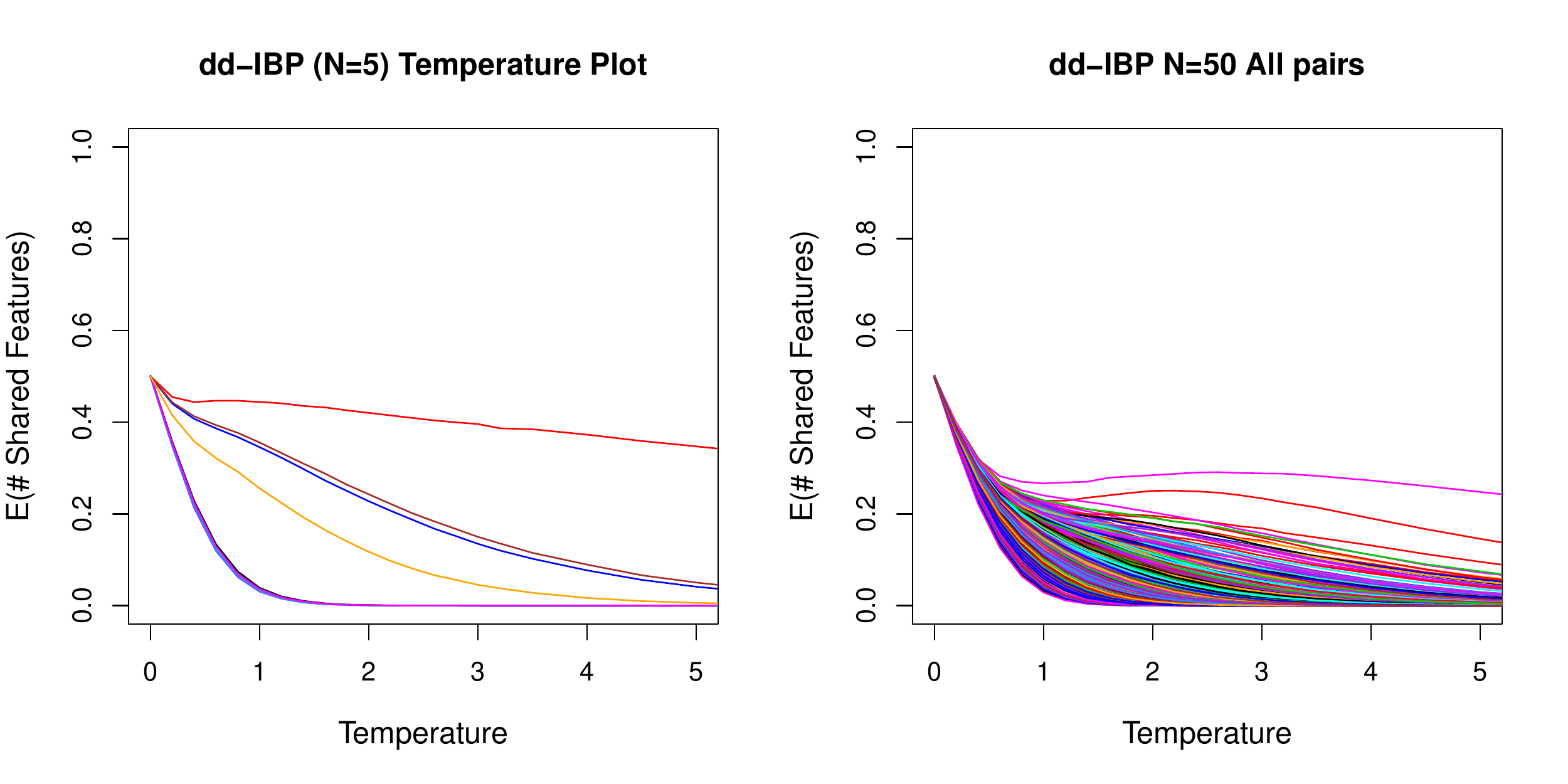}
\caption{A view of the average feature sharing as a function of temperature for
the AIBD and dd-IBP using sample sizes of $N=5$ and $N=50$, where $\alpha =1$.
This figure shows how the AIBD retains the overall average of the expected
number of shared features of the IBP, while on average the dd-IBP with a
sequential proximity matrix shares less, as the temperature increases. For each
plot the permutation of the data, $\bm{\rho }$, was integrated out of the model,
similar to Figure~\ref{temp_plot} but using Monte Carlo integration.}
\label{aibd_N}
\end{figure}

To sample from the AIBD with a greater number of customers, we used pairwise
distance information from the $N=50$ states in the USArrests dataset in
R. To obtain the similarity matrices used, first we centered and scaled
the USArrests dataset. Then, we calculated a pairwise symmetric Euclidean
distance matrix based on all the data. For $N=50$, we simply used all states,
but for $N=5$ we used the same states that were used to calculate Table~\ref{sim_matrices}, which were centered and scaled separately. The simulation
results for the AIBD are shown in the top row of Figure~\ref{aibd_N}.

Note that in the IBP, which corresponds to the AIBD with $\tau =0$, the
expected number of shared features between all customer pairs is
$\alpha /2$. For any $N$, due to exchangeability, the number of shared
features is the same for all customer pairs. Recall that the first customer
samples $X_{1} \sim \text{Poisson}(\alpha )$ dishes. The second customer
will take each dish sampled by customer 1 with probability $1/2$. Thus
the total number of shared features between customers 1 and 2, is
$X_{1,2} \sim \text{Binomial}(X_{1}, 1/2)$. By the law of total expectation,
$E(X_{1,2}) = E(E(X_{1,2}|X_{1})) = E(X_{1}/2) = \alpha /2$. Since any
permutation of customers results in the same probability distribution for
the IBP, the expected number of shared features for the first and second
customer is the same for other pairs. While the AIBD is not exchangeable,
it can be shown by simulation that across any temperature $\tau $ and sample
size $N$, the expected number of shared features averaged across all pairs
is $\alpha /2$. Thus, the AIBD allows each pair to share features differently;
but across all pairs, the mean of the expected number of shared features
is the same as the IBP.

For the AIBD, as customers are added to the process (i.e., as $N$ grows)
the behavior of the average feature sharing changes, as seen in the top
two plots of Figure~\ref{aibd_N}. For high temperatures, a customer pair
can share, on average, more or less than when there are more customers
in the restaurant. At $\tau =5$ and $N=5$, the expected number of shared
features for the 10 possible pairs range from $0.35$ to $0.73$, when
$N=50$ this range increases by roughly 60\% for those same pairs. Thus
increasing $\tau $ and $N$ allows more disparity between the average feature
sharing of customer pairs.

\subsection{Comparison of the AIBD's and the dd-IBP's Properties}
\label{aibd_prior_summary}

Although both use the pairwise distances of items, the AIBD and the dd-IBP
have some notable differences. To compare the AIBD's properties to the
dd-IBP's, we refer to Figure~\ref{aibd_N}. The bottom row of the figure
shows that as the temperature increases in the dd-IBP, on average, pairs
of customers share less. Asymptotically, all average feature sharing goes
to zero; that is, at a temperature of infinity, all customer pairs will
share no features. We find similar behavior for the dd-IBP with a symmetric
proximity matrix; plots for that distribution, similar to Figure~\ref{aibd_N}, can be found in Figure~\ref{symDDIBPFeats} in the supplementary
material. Additional exploration of feature sharing in the AIBD and the
dd-IBP can be found in Section~\ref{Append:FeatSharing} of the supplementary
material.

A table comparing several properties of the AIBD and dd-IBP to the IBP
is shown in Table~\ref{prior_summary_table}. In the dd-IBP, on average
the customers share less as $\tau $ increases. As a result, all lines in
the dd-IBP plots in Figure~\ref{aibd_N} go to zero as
$\tau \rightarrow \infty $. This may be sensible in cases where we want
customers to share less than the IBP. However, it does not seem possible
to allow certain customer pairs to share more than the IBP. In contrast,
the AIBD up-weights or down-weights average sharing depending on the pairwise
similarity value while still having the same average expected number of
shared features across all pairs as the IBP. Larger $N$ increases the disparity
between pairs in the AIBD, but appears to decrease the disparity between
pairs for the dd-IBP.

\begin{table*}
\begin{tabular}{| l | c c |}
\hline
Property & AIBD & dd-IBP \\ \hline
Tractable pmf & Yes & No \\
Reduces to IBP when $\tau =0$ & Yes & In one case\\
E(\# Features) & Same as IBP & Different from IBP \\
E(\# Features per customer) & Same as IBP & Different from IBP \\
E(\# Total shared features) & Same as IBP$^{*}$ & Different from IBP$^{**}$ \\
E(\# Shared features per customer) & Higher or Lower & Different from IBP$^{**}$ \\
\hline
\end{tabular}
\caption{Summary of the properties of the AIBD and dd-IBP to the IBP. Details of
the first two properties can be found in Section~\ref{sec:aibd}, and the
remainder can be found in Section~\ref{sec:aibdprior}, focusing on the dd-IBP's
properties in Section~\ref{aibd_prior_summary}. $*$ Demonstrated via simulation.
$**$ We consistently found this to be lower than the IBP in simulations.}
\label{prior_summary_table}
\end{table*}

\begin{figure}[t!]
\includegraphics[width=\textwidth]{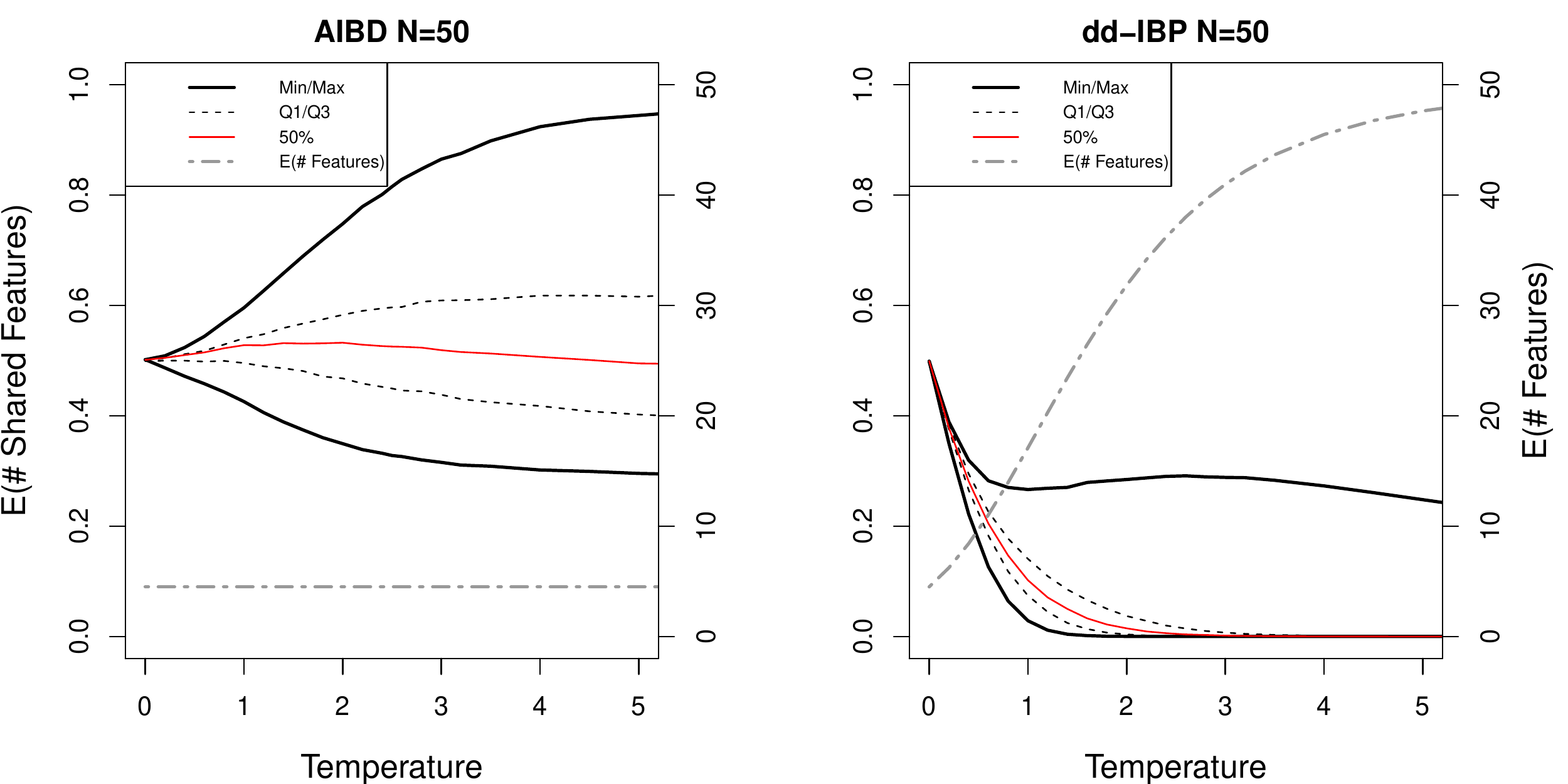}
\caption{The distribution of the total number features overlayed with the
pairwise expected shared feature plots for the AIBD and the dd-IBP for $N=50$.
The Q1 and Q3 lines represent the quartiles of the number of shared features.
This figure highlights that as the temperature increases, the average number of
features (gray dashed line) remain fixed in the AIBD, but grows rapidly in the
dd-IBP. It also shows how the median number of shared features (red line) of the
AIBD and the dd-IBP change as a function of temperature.}
\label{comp_feat}
\end{figure}

For the dd-IBP, a major consequence of the expected number of shared features
going to zero as $\tau \rightarrow \infty $ is that the total number of
features in the distribution increases to $\alpha N$. For $N=50$ and
$\alpha =1$, the expected number of features in the AIBD is the 50th harmonic
number, or 4.5 for all temperatures. For the dd-IBP in this example, however,
the expected number of features can range from 4.5 to 50, depending on
the temperature chosen. This is shown in Figure~\ref{comp_feat}. The dd-IBP
reduces sharing and borrowing of strength between features, which is a
primary reason to use the IBP. This can potentially result in higher computational
burdens when a $\bm{Z}$ from the dd-IBP has a larger number of columns
than the AIBD. One reason why the average number of shared features in
the AIBD does not tend to zero is that the AIBD preserves the distribution
of the total number of features for fixed $\alpha $ and $N$. Thus, in the
AIBD, the distribution of the number of columns is unchanged from the IBP.
The AIBD only influences how features are shared.

The primary point of both the dd-IBP and the AIBD is that two data points
with similar covariates should generally share more features than two with
less-similar covariates. However, our approach to achieving this differs
from the dd-IBP on a key point. In the AIBD, the total number of features
(and, therefore, the total amount of sharing) is controlled with the mass
parameter $\alpha $. It is not influenced by the similarity/proximity matrix.
Further, we also designed the AIBD such that the similarity matrix has
the sole influence on how the total amount of sharing (as determined by
the mass parameter) is divided. For example, the IBP with a particular
value for the mass parameter induces a distribution over the amount of
sharing. The AIBD, with this same value for the mass parameter, also has
the same total amount sharing on average, but the AIBD provides the flexibility
to divide it according to the similarity matrix. In contrast, the dd-IBP
uses the pairwise distance information to influence both the number of
features and the sharing configurations. That is, although both are based
on pairwise distances, the AIBD and dd-IBP use them to achieve fundamentally
different distributions and, together, they provide choice for the statistician.

\subsection{The Similarity Function}
\label{sec:simFunct}

\begin{figure}[t!]
  \begin{tabular}{cc}
    (a) Constant similarity function & (b) Reciprocal similarity function \\
    \includegraphics[scale=.25]{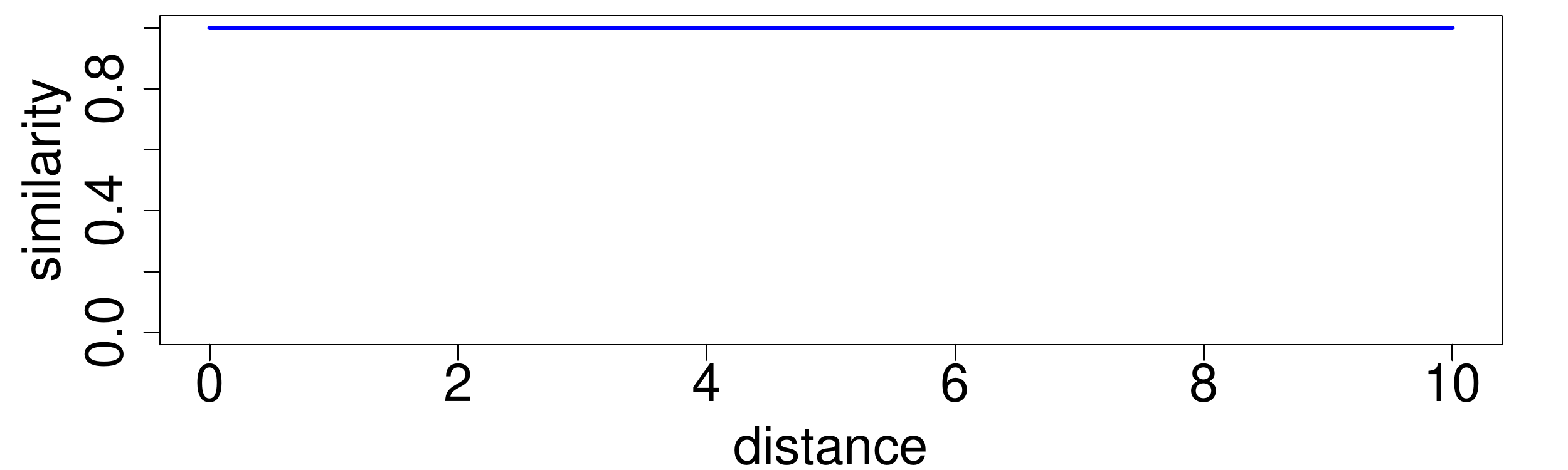} &
    \includegraphics[scale=.25]{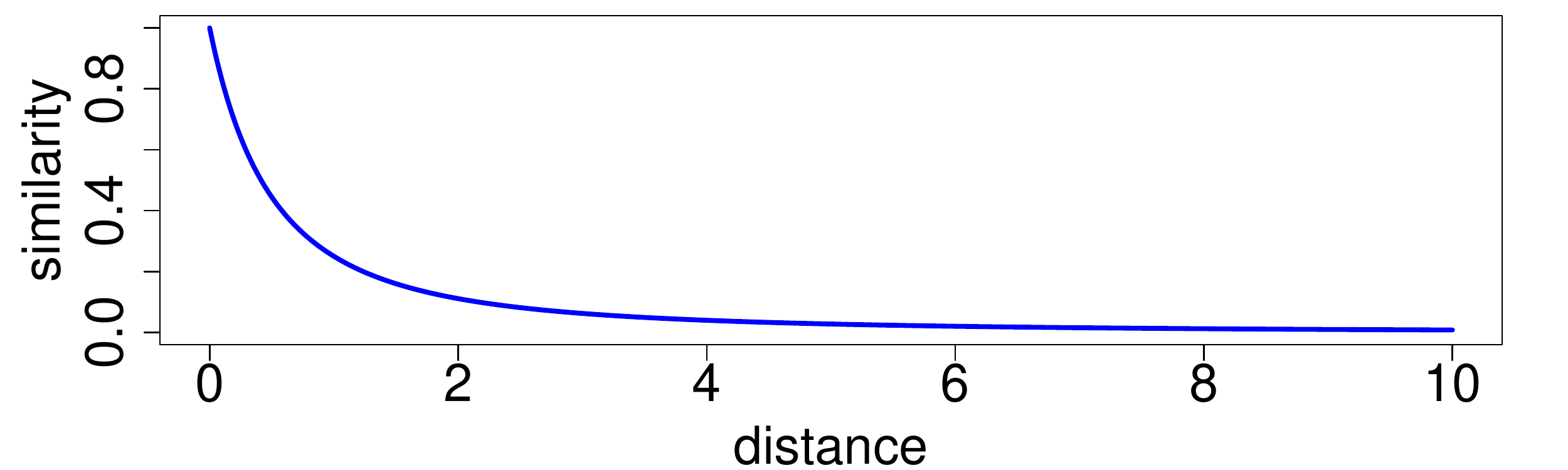} \\
    \includegraphics[scale=.25]{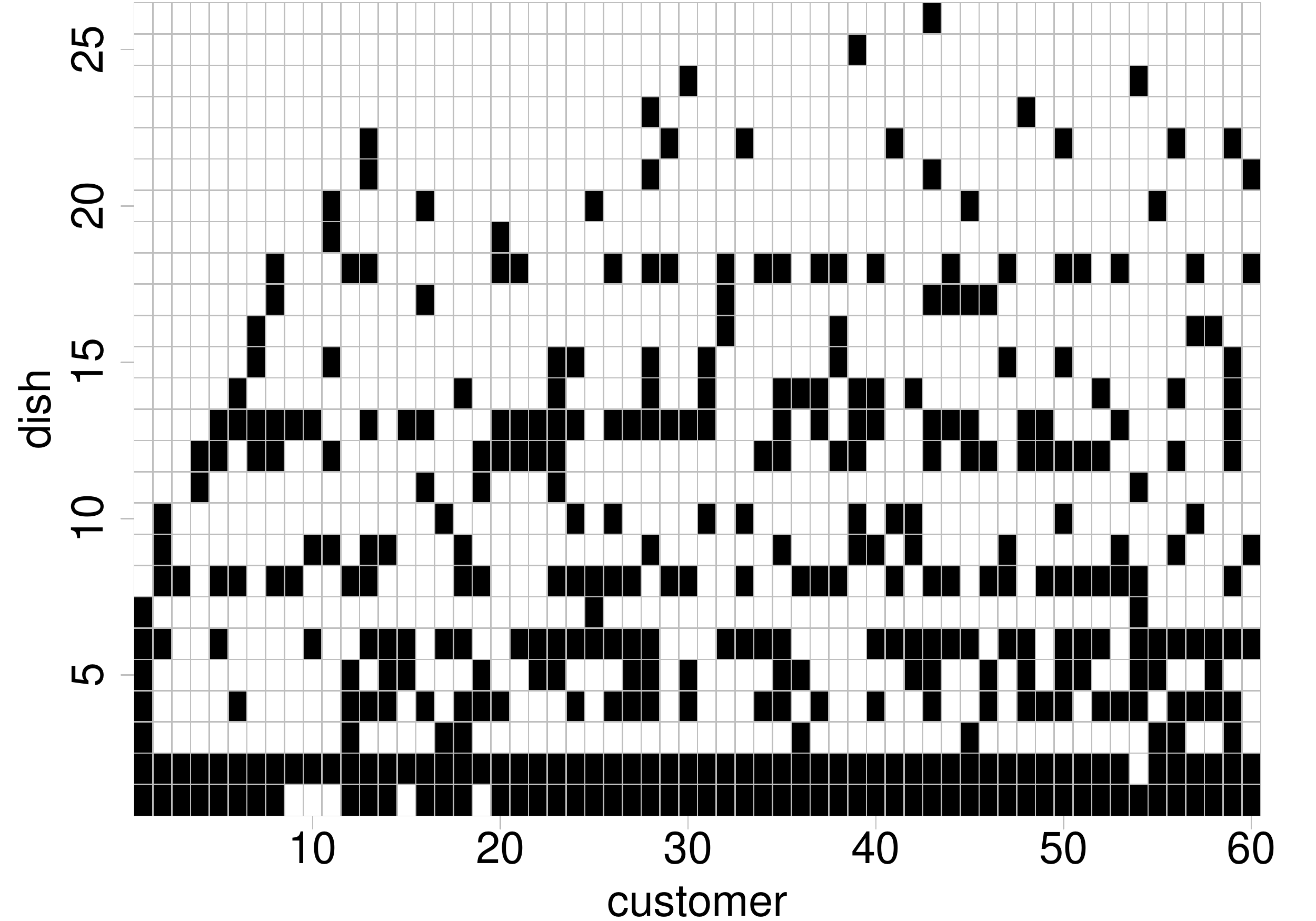} &
    \includegraphics[scale=.25]{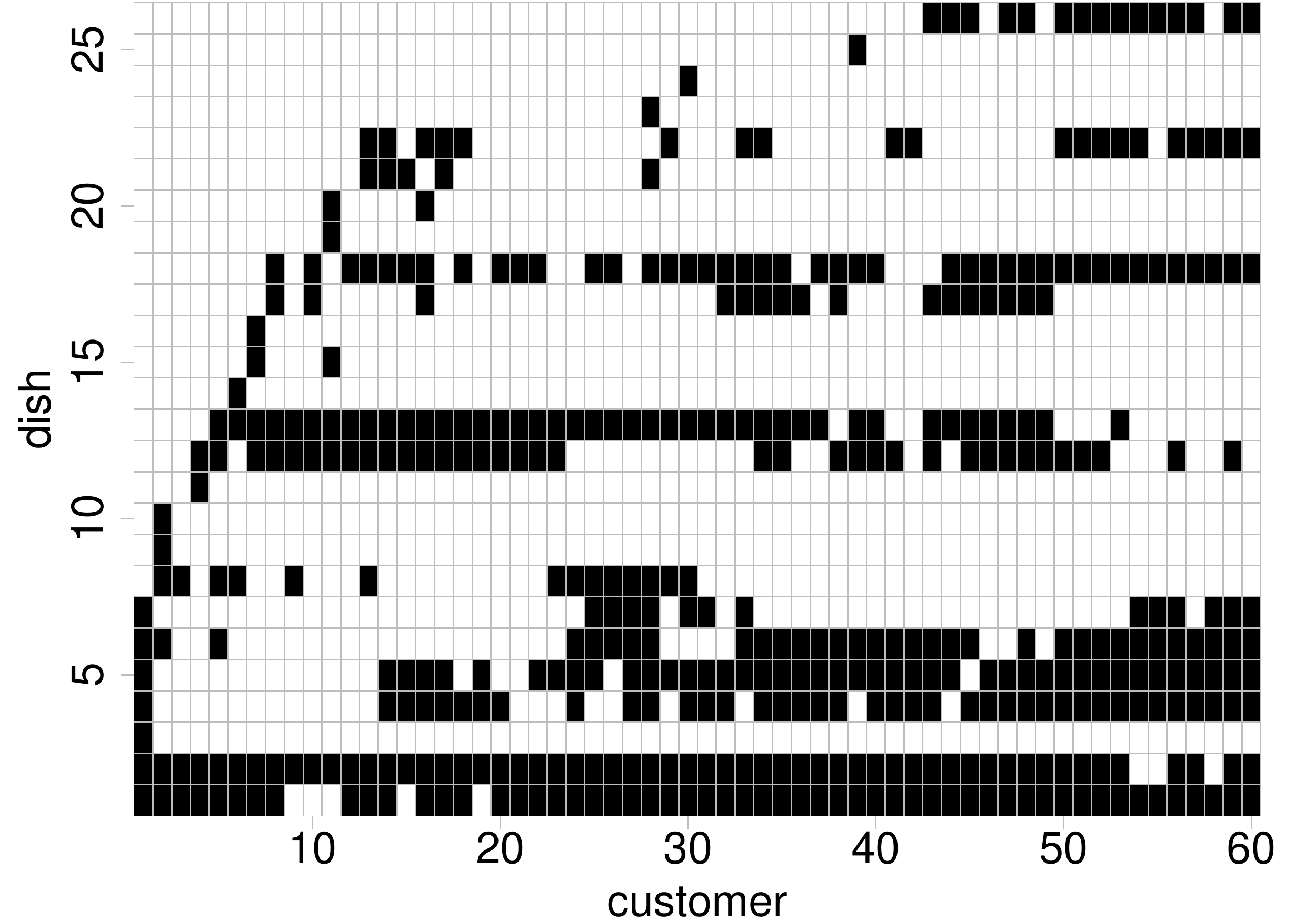} \\
    (c) Exponential similarity function & (d) Window similarity function \\
    \includegraphics[scale=.25]{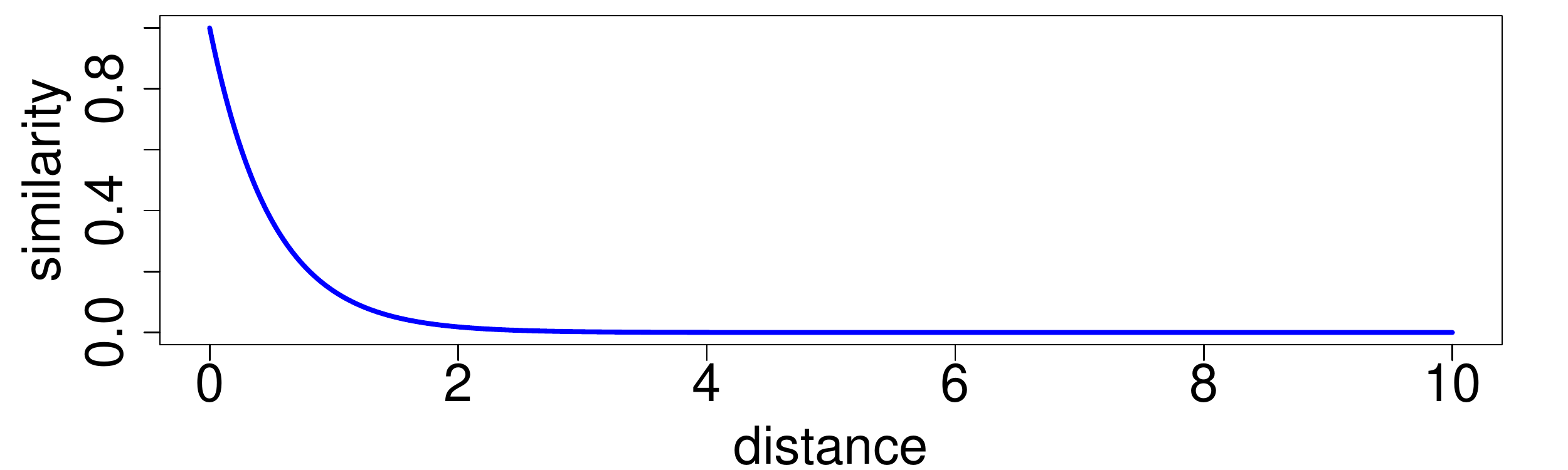} &
    \includegraphics[scale=.25]{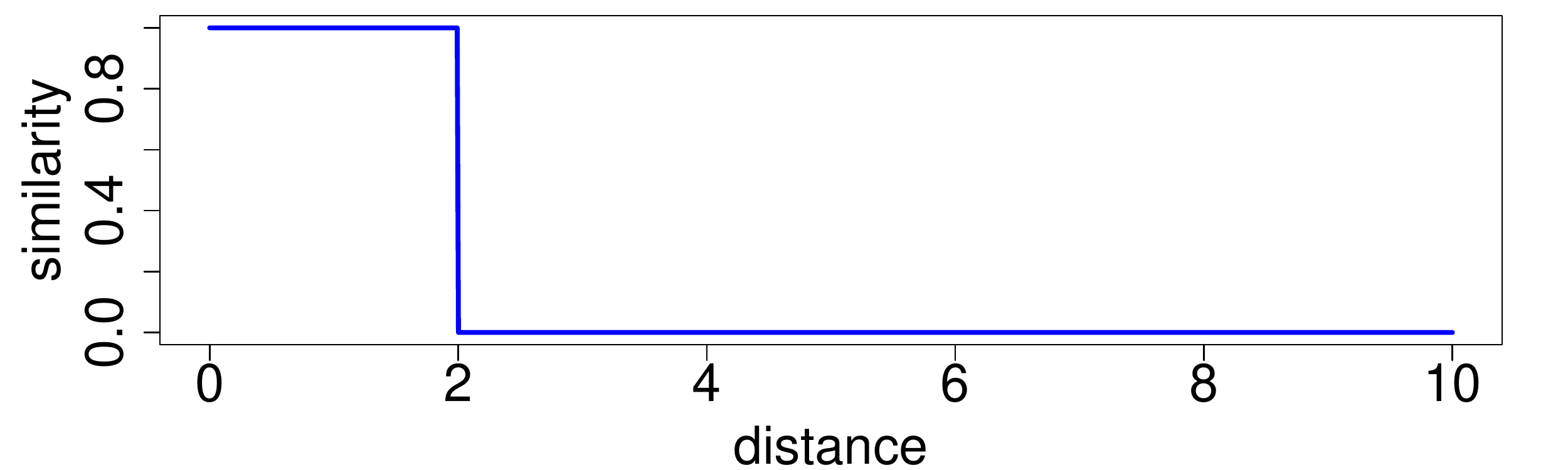} \\
    \includegraphics[scale=.25]{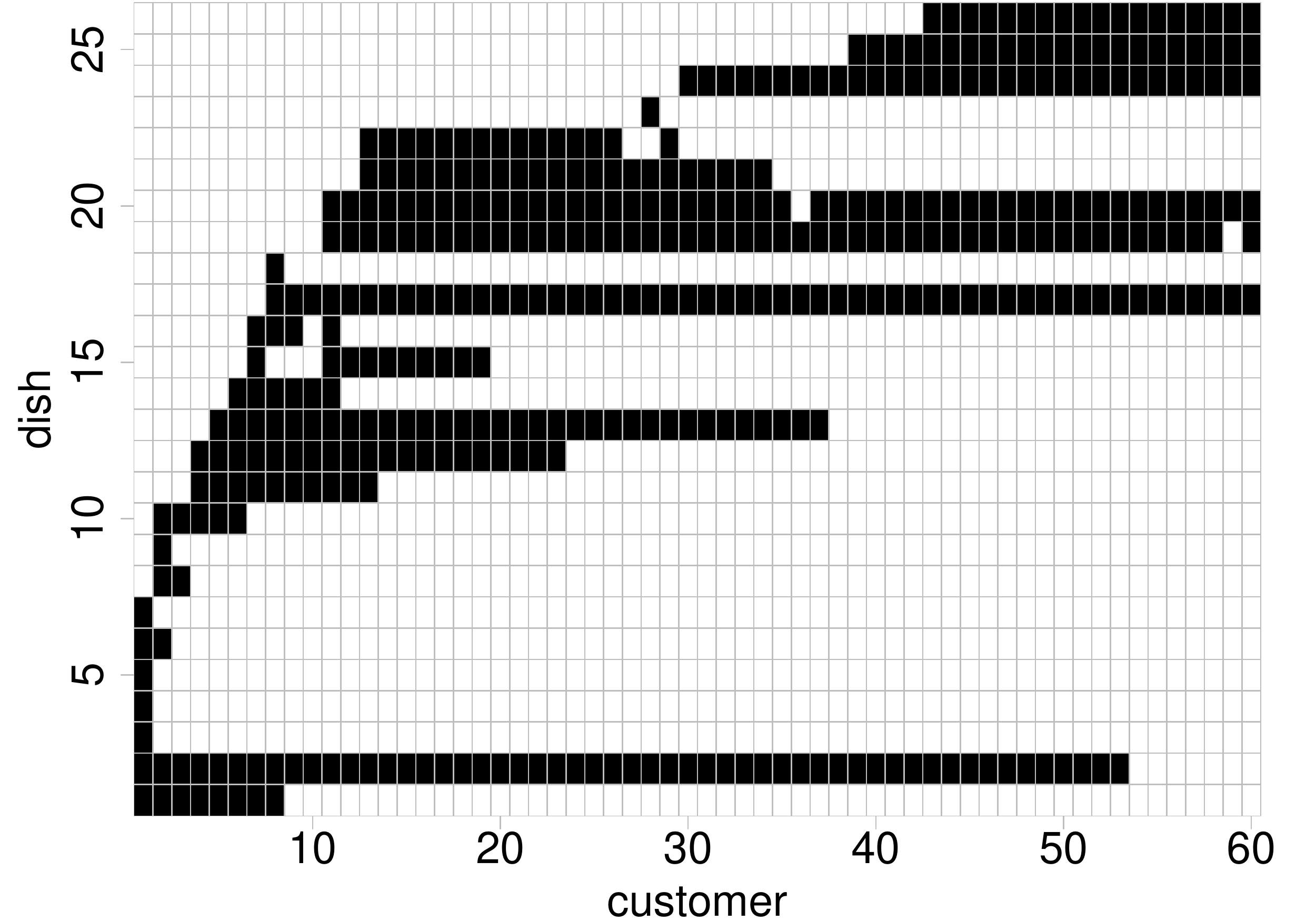} &
    \includegraphics[scale=.25]{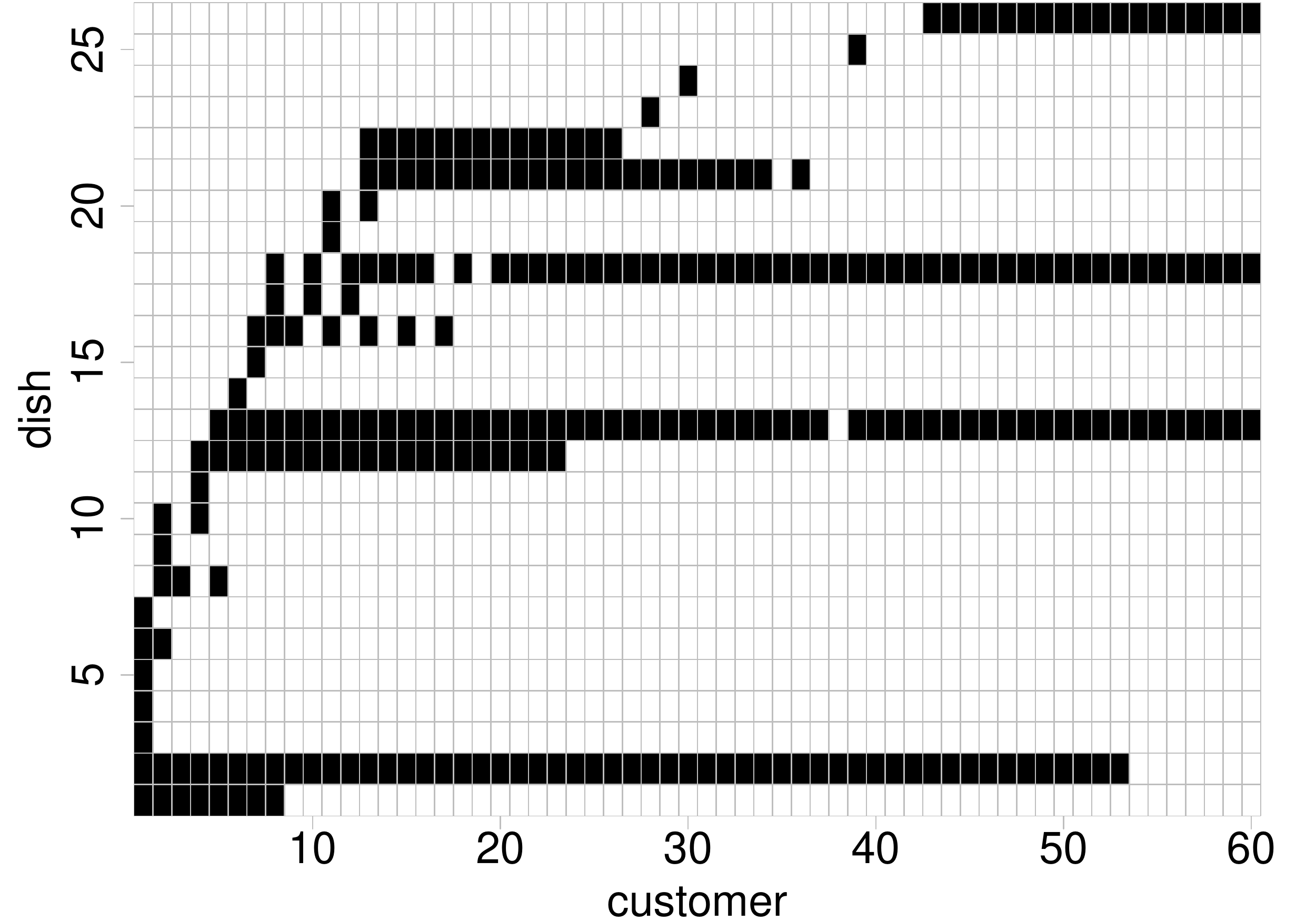}
  \end{tabular}
\caption{A draw from the AIBD under the (a) constant, (b) reciprocal, (c)
exponential, and (d) window functions. For the window similarity, a width of 2
was used; for the reciprocal similarity, a shift of 1 was used. For all
simulations, the number of customers $N$ was 60, the mass $\alpha $ was 5, the
permutation was $\{1,\dots ,N\}$, the temperature $\tau $ was 2, and the
distance used was the absolute temporal distance ($d_{i,j} = |i - j|$).}
\label{fig:decay-functions}\vspace*{-3.5pt}
\end{figure}

The similarity function can be chosen to accommodate a variety of feature-sharing
behaviors. Figure~\ref{fig:decay-functions} shows the transpose of the
feature allocation $\bm{Z}$ simulated under the constant, reciprocal (with
shift$=$1), window (with width$=$2), and exponential functions. A mass of
$\alpha =5$, a temperature of $\tau =2$, and the absolute temporal distance
($d_{i,j} = |i-j|$) for 100 customers were used. The same random number
generator seed was used to generate the figures, so the first two customers
in each scenario take the same dishes. Note that the constant similarity
reduces the AIBD to the IBP.

\begin{figure}
  \begin{tabular}{cc}
    (a) Constant similarity function & (b) Reciprocal similarity function \\
    \includegraphics[scale=.40]{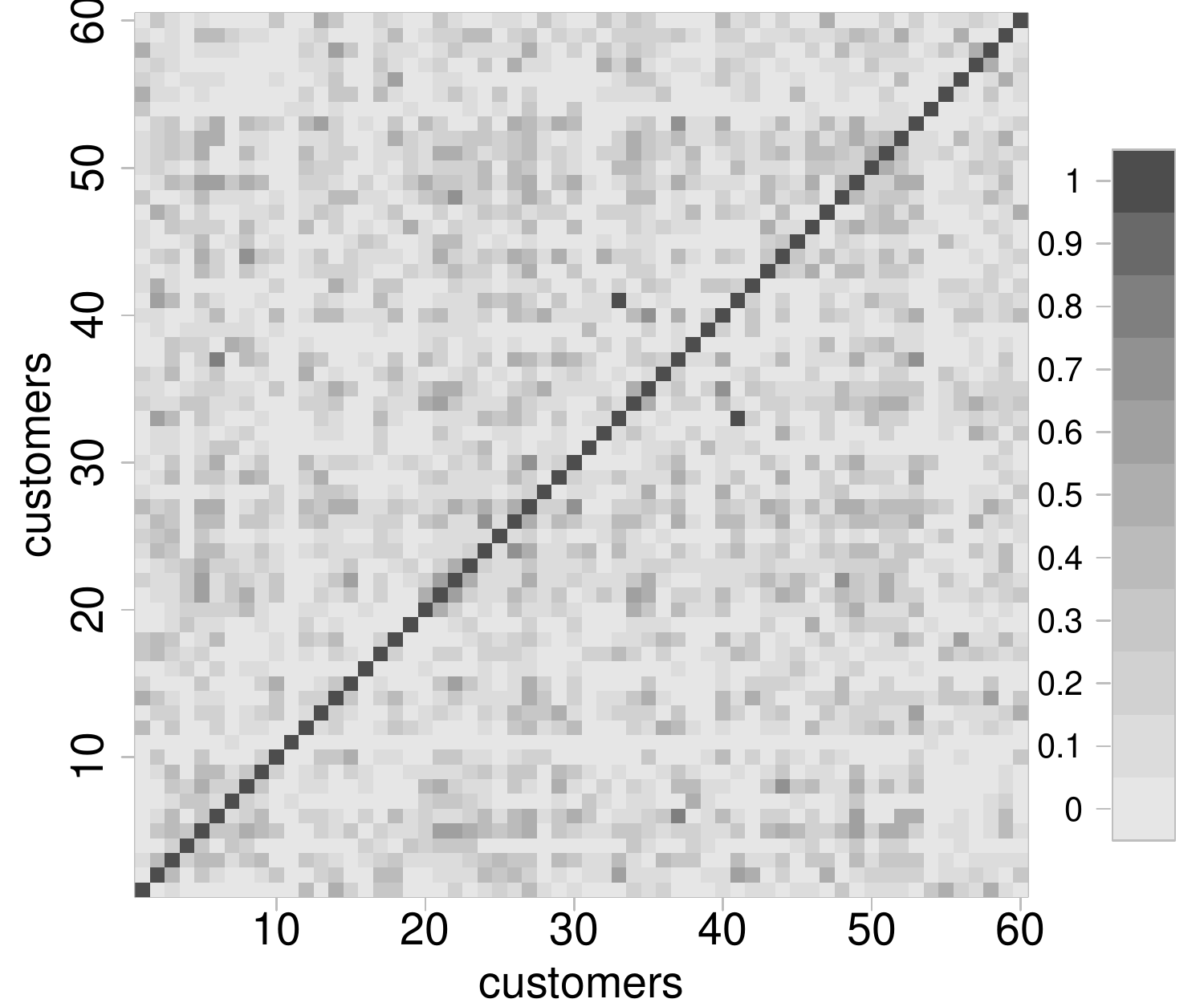} &
	  \includegraphics[scale=.40]{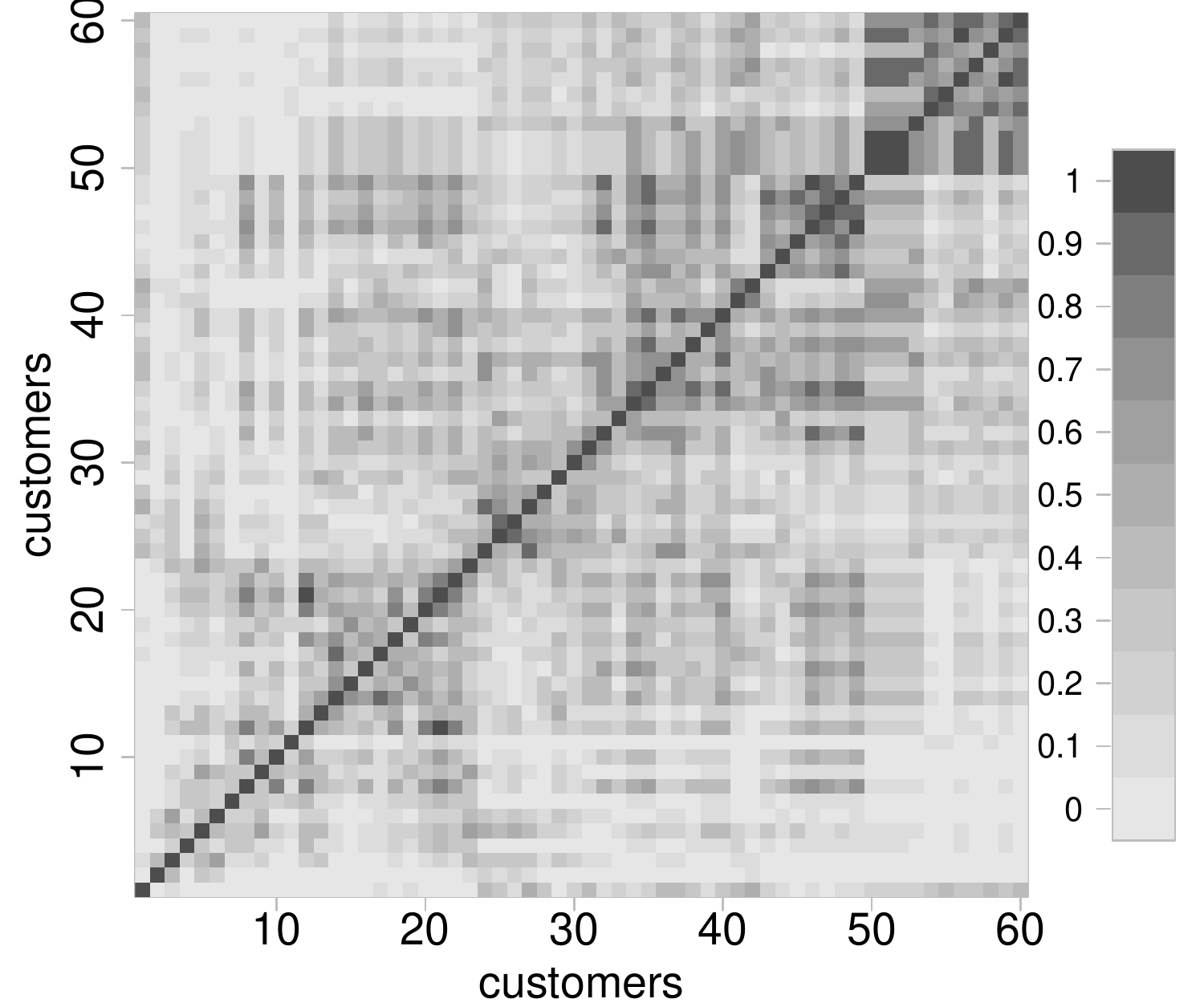} \\
	  \\
    (c) Exponential similarity function & (d) Window similarity function \\
    \includegraphics[scale=.40]{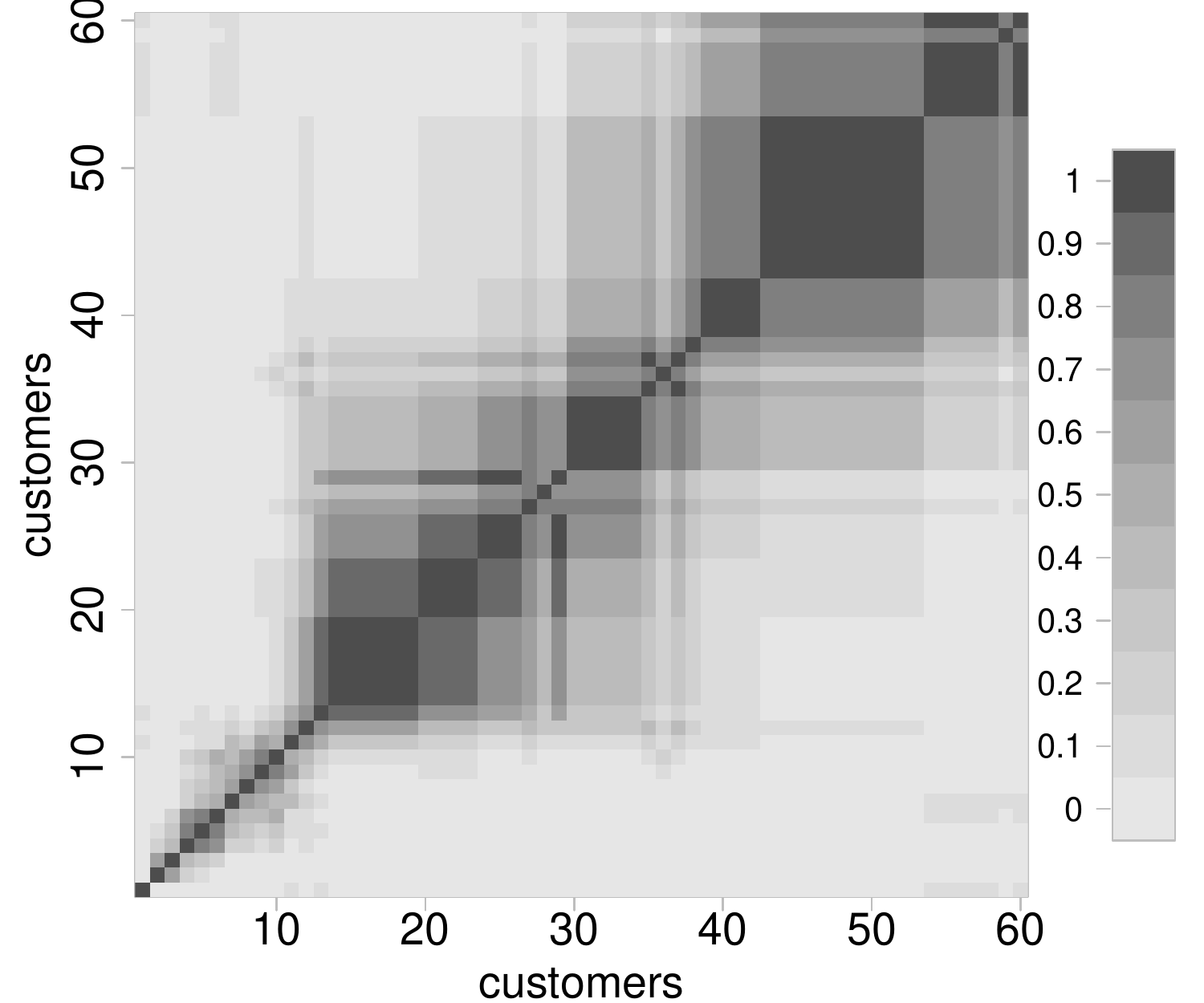} &
    \includegraphics[scale=.40]{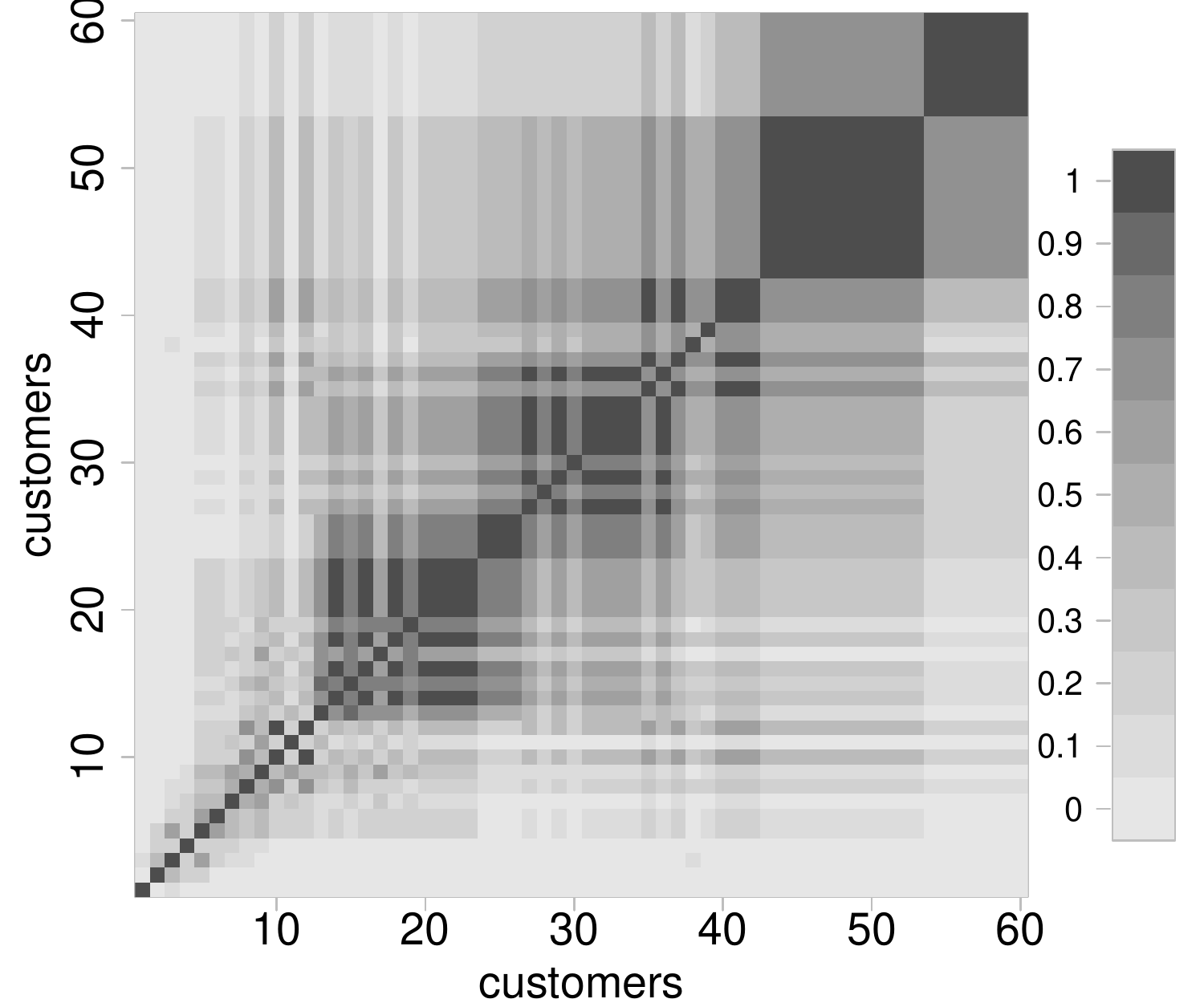} \\
  \end{tabular}
\caption{Squared correlation matrix between customers drawn from the AIBD under
various similarity functions.}
\label{fig:aibd-corr}
\end{figure}

Considering the correlation structures between observations can provide
additional intuition for selecting the similarity functions. To illustrate,
Figure~\ref{fig:aibd-corr} provides a view of the squared correlation between
customers, computed from the feature allocations under the four similarity
functions. Unsurprisingly, under the IBP (constant similarity), correlations
between proximal customers are weak. Under the exponential and window similarity
for the temperature of 2, strong correlations are observed between proximal
customers. Correlations appear to diminish as customers get further apart.
Under the reciprocal similarity with a shift of 1 and temperature of
2, the correlation structure is less pronounced. Moderate correlation can
be observed between proximal customers, but weakens at longer ranges.

\subsection{Lack of Marginal Invariance}
\label{sec:marg-invar}

Unlike the IBP, the AIBD and dd-IBP are not marginal invariant. Both distributions
are such that the distribution defined for $n-1$ customers is not the same
as the distribution obtained by marginalizing over the last customer. Thus,
data analysis with the AIBD and dd-IBP is limited to cases where we are
fully aware of the pairwise distances among all data points. By construction
and as desired, adding another observation changes the relative relationships
among the previous observations. Take, for example, the AIBD similarities
in Table~\ref{sim_matrices}. If New Hampshire was never considered, Iowa
and Wisconsin will share more features because they are relatively closer.
Conversely, having New Hampshire in the model greatly changes the relative
relationships, an effect that persists even when marginalizing over New
Hampshire.

\section{AIBD Posterior Sampling}
\label{post_samp}

We now describe how to sample from the joint posterior distribution of
the model parameters in the LGLFM (i.e., $\bm{Z}$, $\tau $, $\alpha $,
$\bm{\rho }$, $\sigma _{A}$ and $\sigma _{X}$), presented in Section~\ref{sec_lglfm}, with an AIBD prior on the feature allocation. We suggest
priors for the parameters and a Metropolis-Hastings-within-Gibbs sampling
algorithm. Using the likelihood in (\ref{colapLik}), we can write
the full joint posterior as shown in the following equation:
\begin{equation}
p(\bm{Z}, \bm{\rho }, \alpha , \tau , \sigma _{X}, \sigma _{A} |
\bm{X}, \bm{\Lambda }) \propto p(\bm{Z}|\alpha , \bm{\rho },\tau )~p(
\bm{\rho })~p(\alpha )~p(\tau )~ p(\sigma _{X},\sigma _{A})~p(\bm{X}|
\bm{Z}, \sigma _{X}, \sigma _{A}).
\label{full_post}
\end{equation}

\subsection{Posterior Sampling of the Feature Allocation Z}
\label{post_Z}

We use an AIBD prior on $\bm{Z}$ along with prior distributions on the
other parameters in conjunction with the likelihood, and suggest the following
algorithm to sample from the posterior. We emphasize that this algorithm
is valid only for the collapsed likelihood defined in (\ref{colapLik}).
We suppose it could be adapted to other prior distributions on
$\bm{Z}$.

From (\ref{full_post}), the full conditional distribution of
$\bm{Z}$ is proportional to
\begin{equation*}
p(\bm{Z}|\alpha , \bm{\rho }, \tau )~p(\bm{X}| \bm{Z}, \sigma _{X},
\sigma _{A}).
\end{equation*}
Instead of proposing an entirely new $\bm{Z}$ matrix, we update
$\bm{Z}$ row-by-row. Define a singleton feature for row $i$ to be a feature
that only customer $i$ has. In other words, a singleton feature for customer
$i$ is a column in $\bm{Z}$ of all zeros except for a 1 on row $i$. Define
the non-singleton features to be features that are owned by any of the
other customers. For each row $i \in \{1,2,\ldots,N\}$ in $\bm{Z}$:
\begin{enumerate}
\item Let $m_{1}, m_{2}, \ldots m_{h} \in \mathcal{K}$ be the collection of
column indices of the non-singleton features in the current state of
$\bm{Z}$, where $h$ is the total number of non-singletons. If $h=0$, skip
to step 3. If $h>0$, generate a random permutation of the collection of
indices in $\mathcal{K}$. We update the non-singleton columns in this order
in step 2.
\item Start by updating the non-singleton features on row $i$ one at a
time using the Metropolis-Hastings algorithm in the permuted order as generated
in step 1. Denote $z_{i,m}$ to be the binary number in the $i^{th}$ row
and $m^{th}$ column of $\bm{Z}$. We update each element $z_{i,m}$ for each
$m \in \mathcal{K}$ sequentially according to the permutation of
$\mathcal{K}$. For each $m \in \mathcal{K}$:
\begin{enumerate}
\item Define the active feature to be the feature in the $m^{th}$ column
of the current state of $\bm{Z}$. This is the feature that is currently
being updated.
\item Propose $z^{*}_{i,m} = 1 - z_{i,m}$ (i.e., the opposite of the current
state of $z_{i,m}$). Let $\bm{Z^{*}}$ be the same as $\bm{Z}$, except for
$z^{*}_{i,m}$ in place of $z_{i,m}$.
\item Since the order of the columns does not matter, different updates
may result in the same proposed feature allocation. Thus, the proposal
distribution is not symmetric. The Hastings ratio is computed by dividing
the number of features identical to the active feature $d^{*}$ in
$\bm{Z}^{*}$ by the number of features identical to the active feature
$d$ in $\bm{Z}$. The active feature is also counted in this total, so if
the active feature is distinct in both the current and proposed states,
the ratio is 1. From the example in Figure~\ref{Z_update}, $m=4$ and
$h=4$, and the active feature is column 4. In the current state, there
are $d=2$ features identical to column 4 (including itself), and
$d^{*}=3$ features identical to column 4 in the proposed state.
\item Compute the Metropolis-Hastings ratio
$MHR =
\frac{p(\bm{Z^{*}}|\alpha , \bm{\rho }, \tau )p(\bm{X}| \bm{Z}^{*}, \sigma _{X}, \sigma _{A})}{p(\bm{Z}|\alpha , \bm{\rho }, \tau )p(\bm{X}| \bm{Z}, \sigma _{X}, \sigma _{A})}
\frac{d^{*}}{d}$ and update $z_{i,m}$ to $z_{i,m}^{*}$ with probability
min(1, $MHR$), else leave $z_{i,m}$ unchanged.
\end{enumerate}
\item Now we propose new singleton features for customer $i$.
\begin{enumerate}
\item We first evaluate the unnormalized probability mass of the full conditional
distribution of $\bm{Z}$ after adding $0, 1, 2, \ldots $ features (with
all other parameters and rows in $\bm{Z}$ held constant). Since we cannot
check a theoretically infinite number of features, we stop considering
new features once we obtain a mass that is less than a specific fraction
(e.g., we used $1/1000$) of the highest mass. This should cover most reasonable
posterior values and the fraction is a tuning parameter that can be adjusted
if desired. This truncation step makes this algorithm an approximate sampler,
but it can closely mimic an exact sampler. See Section~\ref{MCMC_Simulation} in the supplementary material for a simulation study
on the accuracy of this sampling algorithm.
\item We estimate probabilities $p_{0}, p_{1}, p_{2}, \ldots $ of adding 0,1,2,\ldots
singleton features by dividing each unnormalized mass in the previous step
by the sum of all masses. Add $j$ singletons to customer $i$ with probability
$p_{j}$.
\end{enumerate}
\end{enumerate}

\begin{figure}
\begin{minipage}[t]{.49\textwidth}
\begin{center}
Current State $\bm{Z}$

$
\begin{bmatrix}
1 & 0 & 1 & \underline{0} & 1
\\
1 & 1 & 0 & 1 & 1
\\
0 & 0 & 0 & 0 & 0
\\
\end{bmatrix}
$
\end{center}
\end{minipage}
\begin{minipage}[t]{.49\textwidth}
\begin{center}
Proposed State $\bm{Z}^{*}$

$
\begin{bmatrix}
1 & 0 & 1 & \underline{1} & 1
\\
1 & 1 & 0 & 1 & 1
\\
0 & 0 & 0 & 0 & 0
\\
\end{bmatrix}
$
\end{center}
\end{minipage}

\caption{Example of a posterior update from step 2. Customer 1 has one singleton
feature (column 3) and four non-singleton features. For customer 1, we will
individually update the columns 1,2,4, and 5 in a random order. The underlined
number $z_{14}$ is being updated and thus the active feature is column 4. There
are 3 features identical to the active feature (including itself) in the
proposed state and 2 identical features in the current state. The Hastings ratio
is 3/2 to account for the asymmetric proposal.}
\label{Z_update}
\end{figure}

After going through all rows, the result is one scan of the Markov chain
updates for $\bm{Z}$. We then sample from the other parameters, which we
describe in the next section.

\subsection{Sampling the Other Parameters}
\label{post_other}

After updating $\bm{Z}$, we proceed to update the other parameters ($
\alpha $, $\bm{\rho }$, $\tau $, $\sigma _{X}$, and $\sigma _{A}$). The
parameters $\alpha $ and $\tau $ are sampled univariately; while
$(\sigma _{X}, \sigma _{A})$ is sampled jointly, and $\bm{\rho }$ is sampled
as a vector.

For the mass parameter $\alpha $, we suggest using a gamma prior because
it is conditionally conjugate. If a Gamma($a_{\alpha }$, $b_{\alpha }$) prior
(with expectation $a_{\alpha }/b_{\alpha }$) is used, then the resulting
conditional posterior is a draw from a Gamma($a_{\alpha } + T_{\bm{Z}}$,
$b_{\alpha } + H_{N}$). $T_{\bm{Z}}$ indicates the total number of features
in the current state of $\bm{Z}$ and $H_{N}$ is the $N^{th}$ harmonic number.

For the permutation parameter $\bm{\rho }$, we suggest using a discrete
uniform on all possible permutations of the integers 1 through N, unless
there is natural ordering in the data. Using the discrete uniform prior,
we update $\bm{\rho }$ with a random walk using a discrete uniform proposal.
For small N, this can be done by sampling from any of the $N!$ possible
permutations. However, for larger $N$, this could lead to low acceptance
rates. As such, we recommend only updating $k_{\rho }$ of the elements in
the permutation at a time, where $k_{\rho }$ is a tuning parameter that
controls how quickly the permutation space is explored. In our experience
choosing $k_{\rho }$ to obtain an acceptance rate of around 25\% produced
good mixing. As $N$ gets larger, $k_{\rho }$ may need to stay at a fixed
quantity to get good acceptance rates. The steps to sample a new
$\bm{\rho }$ are outlined as follows. First randomly select
$k_{\rho }$ indices in the permutation to update. Next, randomly shuffle
the $k_{\rho }$ indices, while leaving the other $N-k_{\rho }$ indices fixed,
to generate a proposed permutation. Finally, calculate the Metropolis acceptance
ratio, $R_{\rho }$ and accept the proposed permutation with probability
min(1, $R_{\rho }$). Checking for good mixing and convergence for
$\bm{\rho }$ is more nuanced than for the other parameters. We provide some
recommendations in Section~\ref{Sec:app-consider} to assess the convergence
for $\bm{\rho }$.

The temperature parameter $\tau $ in the likelihood does not appear to
have a conjugate prior. Therefore, any prior with positive continuous support
might be reasonable. For our implementation, we choose a Gamma prior. To
draw from the conditional posterior of $\tau $ we use a Metropolis step
with a Gaussian random walk proposal and reject proposals outside the support
of $\tau $.

For the variance components of the likelihood, we recommend using any positive
continuous prior, such as a Gamma prior on $\sigma _{X} $ and
$\sigma _{A}$. Since $\sigma _{X}$ and $\sigma _{A}$ are typically negatively
correlated in the posterior, we used a bivariate Gaussian random walk to
update both parameters simultaneously, again rejecting proposals outside
the support of $(\sigma _{A}, \sigma _{X})$.

Due to the computational burden of updating $\bm{Z}$ relative to updating
the other parameters, we recommend updating the other parameters several
times for every update of $\bm{Z}$. We updated other parameters 10 times
for every update of $\bm{Z}$ in the application in Section~\ref{dataAnalysis}
to reduce the autocorrelation within the posterior draws, at a negligible
computational cost.

The methods suggested in this section are implemented in the
\textit{samplePosteriorLGLFM} function of the \textit{aibd} R package. The
posterior sampling algorithm can be applied to both the AIBD and, since
it is a special case of the AIBD, the IBP. From our experience, the results
are accurate and the only source of bias comes from the truncation step.
This truncation error can be controlled and appears to be negligible compared
to Monte Carlo error.\vspace*{-2pt}

\section{Data Analysis}
\label{dataAnalysis}\vspace*{-2pt}

In this section, brain measurement data is analyzed using a latent feature
model in several analyses. The LGLFM likelihood, introduced in
\cite{ghahramani2006infinite} and discussed in Section~\ref{sec_lglfm} of this paper, remains the same over each analysis, however,
the prior distributions vary. Three fundamentally different prior distributions
are considered. Specifically, the proposed AIBD, the dd-IBP and the standard
IBP are used as prior distributions on the matrix $\bm{Z}$ and the posterior
results are then compared.\vspace*{-2pt}

\subsection{The Data}\vspace*{-2pt}

The data for these analyses were obtained from a neuroimaging study of
the brains of healthy and Alzheimer's-diseased subjects (see
\citealp{dinov2009efficient}). The data and some details of the study are
available on UCLA's Statistics Online Computational Resource (SOCR) data
page \cite{datawebpage}. Of that data, we consider the 27 Alzheimer's diseased
and 35 normal control subjects. In the study, 56 distinct regions of interest
(ROIs) in the brain are observed; each region has four different measurements.
The four measurements are the surface area (SA), shape index (SI), curvedness
(CV), and fractal dimension (FD). One of the properties of the LGLFM is
that the error terms for each of these measurements are assumed to have
constant variance. Therefore, before modeling, the ROI measurements are
centered and scaled (so each has a mean of zero and a standard deviation
of one). Additionally, the SA measurements are somewhat skewed, thus a
log transformation was performed before centering and scaling those measurements.\vspace*{-2pt}

\subsection{The Analysis}\vspace*{-2pt}

The analysis of this data mirrors the steps taken in
\cite{gershman2014distance}. First, patient age is included in the AIBD
and dd-IBP priors as a distance between patients. Including this distance
makes the both the AIBD and dd-IBP priors nonexchangeable, with the hope
that this extra information will improve the model's predictive performance.
The distances between patients are defined in both priors using the exponential
decay function (i.e.,
$\exp \{ - \text{temperature} \times |\text{age difference}| \}$).\footnote{In
the AIBD prior, 0.00001 is added to each age difference to ensure no two
patients have a distance of 0.} As in their analysis, we use a sequential
proximity matrix for the dd-IBP and fix the order of the patients from
youngest to oldest in both the AIBD and the dd-IBP. The temperatures for
the AIBD and dd-IBP are set at 5 levels (0.4, 0.8, 1.2, 1.6, and 2.0).
Since the IBP prior is exchangeable, it does not include any distance information
between patients. In each model a gamma$(1,1)$ is used as the prior distribution
for the mass parameter (the same as in the dd-IBP code by
\citealp{ddIBPcode}). The LGLFM likelihood is used as defined previously,
with the data being the 4 different measurements in 56 regions of the brain.
The data, $\bm{X}$, are contained in a 62 $\times $ 224 matrix, which is
an over parameterized model unless some type of regularization or dimension
reduction is used.\vadjust{\vspace*{-3.5pt}\goodbreak}

Prior distributions for the variance components of the likelihood also
need to be specified. Since the data are centered and scaled, a reasonable
maximum value for $\sigma _{A}$ and $\sigma _{X}$ is 1. Therefore, in all
models a standard uniform prior is placed on these parameters; i.e.,
$p(\sigma _{A}) = \text{I}\,(0 < \sigma _{A} <1)$ and
$p(\sigma _{X}) = \text{I}\,(0 < \sigma _{X} <1)$.

With the models fully specified, we follow similar steps of the analysis
in \cite{gershman2014distance}. First, we obtain 3,000 MCMC posterior samples
from each model. The first half of the samples is discarded for burn-in.
Each model had 50 MCMC chains, for a total of 75,000 posterior samples.
For each sample, the subjects are randomly divided into training and testing
sets. Using the latent features from the posterior sample of
$\bm{Z}$ as predictors, an L$_{2}$-regularized logistic regression is performed
on the training set (to classify which patients do or do not have Alzheimer's
disease). For the logistic regression, we use the \textit{penalized} function
in the \textit{penalized R} package \citep{penal1,penal2} with the regularization
constant set to $10^{-6}$. Using the results from the logistic regression
model, the test group subjects are then classified to assess performance.
Finally, for each test group's classification, the area under the receiver
operating characteristic curve (AUC) is calculated. Thus we obtain 75,000
posterior samples of AUC, which are used to compare each model's efficacy.

Although the steps in this analysis are the same as in
\cite{gershman2014distance}, based on the information provided in that
article, a few aspects of the analysis cannot be replicated. First, in
the dd-IBP paper a few observations were randomly removed in the classification
to balance the training and testing sets. However, the removed observations
were not identified. Therefore, we do not ignore any observations; i.e.,
the training and testing sets are disjoint but include all 62 subjects.
Also, it is not specified which observations are assigned to the training
and testing sets for classification. In our implementation, we randomly
assign 14 diseased and 18 healthy subjects to the training set, and then
assign the remaining 13 diseased and 17 healthy subjects to the testing
set; this is repeated for each posterior sample.

In total there are five AIBD priors (one for each temperature), five dd-IBP
priors, and one IBP prior. In each of those 11 models we obtained 75,000
posterior samples; the $i^{\text{th}}$ sample of each chain are compared
using identical training and testing sets during classification. For posterior
simulation, the code available at \cite{ddIBPcode} is used for the dd-IBP,
and the functions included in the \textit{aibd} package
\citep{AIBDpackage} are used for both the AIBD and the IBP. Posterior means
and standard deviations of the parameters from the several models are included
in Section~\ref{Append:PostSums} of the supplementary material. One possible
confounding factor in our comparison of the AIBD and the dd-IBP is that
the posterior simulation for the dd-IBP is an approximate MCMC scheme,
as noted in \cite{ddIBPcode}.

\subsection{Comparison Between the AIBD and the dd-IBP}
\label{sec:aibdVddibp}

Each model's performance is measured by how well it correctly classifies
subjects into ``healthy'' or ``diseased''. This performance can be quantified
using the AUC, which ranges between zero and one, with higher values indicating
a better classifier. Using the AUC, from the classifications on the testing
sets, the results of the models with AIBD and dd-IBP priors\vadjust{\goodbreak} are compared.
95\% Monte Carlo confidence intervals for the expected posterior AUC from
the models with AIBD and dd-IBP priors are reported in Table~\ref{aibdANDddibp}. This table shows that, on average, the AIBD has higher
AUC than the dd-IBP at every temperature setting. In Table~\ref{aibd-ddibp-dic} we also include the deviance information criterion
(DIC) scores (as defined by \cite{gelman2013bayesian} and use the penalty
term in their Equation 7.8), for each model to determine which has a better
fit. The DIC indicates that the AIBD is performing better than the dd-IBP.

\begin{table*}
{\small
\begin{tabular}{| l | c c c c c |}
\hline
Temp & 0.4 & 0.8 & 1.2 & 1.6 & 2.0 \\ \hline
AIBD & (0.747, 0.748) & (0.747, 0.748) & (0.750, 0.752) & (0.754, 0.755) & (0.753, 0.754) \\
dd-IBP & (0.677, 0.679) & (0.715, 0.716) & (0.708, 0.710) & (0.719, 0.721) & (0.720, 0.721) \\ \hline
\end{tabular}}
\caption{Posterior Expected AUC. 95\protect \% Monte Carlo confidence intervals
for the expected posterior AUC for the five AIBD and dd-IBP models (with fixed
temperatures and permutation), higher values indicate better model performance.}
\label{aibdANDddibp}
\end{table*}

\begin{table*}
\begin{tabular}{| l | c c c c c |}
\hline
Temp & 0.4 & 0.8 & 1.2 & 1.6 & 2.0 \\ \hline
AIBD & 32531.58 & 32523.37 & 32557.66 & 32513.63 & 32489.31 \\
dd-IBP & 35322.27 & 35005.67 & 35286.97 & 35316.80 & 35144.78 \\ \hline
\end{tabular}
\caption{DIC values for the five AIBD and dd-IBP models (with fixed temperatures
and permutation), lower values indicate better model fit.}
\label{aibd-ddibp-dic}\vspace*{-3.5pt}
\end{table*}

While the AIBD performs better than the dd-IBP under these conditions,
it has the added benefit of easily accommodating a prior on the temperature
parameter. Finding a suitable static value for the temperature parameters
seems neither intuitive nor straight-forward. Allowing prior distributions
to incorporate some amount of uncertainty is arguably more appropriate.
An additional model with an AIBD prior was fit (with 75,000 posterior samples)
with a gamma$(1,1)$ prior distribution set on the temperature parameter
and a uniform distribution on the permutation, $\bm{\rho }$. When taking
advantage of this flexibility the AIBD performs even better, with a 95\%
Monte Carlo confidence interval on the expected posterior AUC of
$(0.758, 0.760)$, and a DIC of 32477. It should be noted that it is also
possible to set a prior on the permutation for the dd-IBP, but the available
software does not have that option. This example demonstrates that the
AIBD prior is better able to capture the distance information (age) between
subjects. On average, the logistic regression trained on each posterior
$\bm{Z}$, learned from the AIBD model when compared to the dd-IBP model,
more accurately classifies subjects as ``healthy'' or ``diseased.''

To further compare the AIBD and the dd-IBP, the expected number of posterior
features in $\bm{Z}$ can be found in Table~\ref{NumFeatures}. These numbers
demonstrate a substantial reduction in dimension from the original 224
data columns plus the age variable. This reduction in dimension is really
larger than it first appears because the entries in feature allocation
matrices are binary, zeros and ones, whereas the data contained in
$\bm{X}$ are continuous values. To determine how much information from
the data is lost in the dimension reduction, we compare the AUC results
from two alternative models. These alternative models are penalized logistic
regressions fit to the same training sets of the centered and scaled data
as the AIBD and dd-IBP models, one includes $\bm{X}$ as the predictors,
and the other includes both $\bm{X}$ and the age of each patient. The average
AUC of these models are 0.778 and 0.785 respectively. So little information
is lost in the dimension reduction and the AIBD's posterior feature allocation
can comprise nearly all the information about a patient's disease state,
while drastically decreasing the dimensionality. This reduction of dimension
is most useful if the results can be interpreted.

\begin{table*}
\begin{tabular}{| l | c c c c c |}
\hline
Temp & 0.4 & 0.8 & 1.2 & 1.6 & 2.0 \\ \hline
AIBD & 68 (61,76) & 68 (61,76) & 66 (60,73) & 66 (59,73) & 65 (59,72) \\
dd-IBP & 87 (56,128) & 104 (74,134) & 95 (68,135) & 105 (76,200) & 105 (73,199) \\ \hline
\end{tabular}
\caption{The average number of features in the posterior samples of $\bm{Z}$ for
the AIBD and the dd-IBP at five different temperatures. Additionally, for each
case, a 95\protect \% posterior probability interval for the number of features
is shown in parentheses.}
\label{NumFeatures}
\end{table*}

\begin{table*}
\begin{tabular}{| l | c c c c c |}
\hline
Temp & 0.4 & 0.8 & 1.2 & 1.6 & 2.0 \\ \hline
AIBD & 4.3 (3.9,4.9) & 4.5 (4.1,5.2) & 4.7 (4.1,5.2) & 4.8 (4.2,5.3) & 4.9 (4.3,5.6) \\
dd-IBP & 4.2 (3.6,4.8) & 3.6 (3.1,4.1) & 3.2 (2.9,3.7) & 2.9 (2.5,3.3) & 2.7 (2.3,3.1) \\ \hline
\end{tabular}
\caption{The average number of customers per feature in the posterior samples of
$\bm{Z}$ for the AIBD and the dd-IBP at five different temperatures.
Additionally, for each case, a 95\protect \% posterior probability interval for
the number of customers per feature is shown in parentheses.}
\label{CustPerFeats}\vspace*{-5pt}
\end{table*}

As discovered in the prior exploration of the AIBD and the dd-IBP, we also
see similar feature sharing trends in the posterior. Table~\ref{CustPerFeats} shows the average number of customers per feature in
the posterior samples for each of the 10 models. These numbers indicate
that the amount of feature sharing goes down in the posterior of the dd-IBP
as the temperature parameter increases. While we see an increase in the
amount of sharing in the AIBD for the same temperatures.

A very rough speed comparison of the posterior sampling algorithms for
the dd-IBP and the AIBD showed that the dd-IBP could do one update of all
parameters in roughly 13 seconds (using $\sim $25 cores), while the AIBD
took approximately 70 seconds (using one core). There are a few issues
that make a full comparison quite challenging. One is that the methods
are implemented in different software; the dd-IBP takes advantage of MATLAB's
multi-threading capabilities, while the AIBD implementation is single-threaded.
Additionally, each MCMC scan is highly dependent on the number of active
features in the current iteration, and the dd-IBP had a higher average
number of posterior features in $\bm{Z}$. The AIBD also gains a time advantage
using compiled code.

\subsection{Comparison Between the AIBD and the IBP}

As demonstrated, the AIBD competes favorably with the dd-IBP as an effective
nonexchangeable prior for latent feature models. In this application we
also want to show that using a nonexchangeable prior can produce better
results than when using an exchangeable prior, specifically the IBP. In
this case the AIBD has the added advantage over the IBP of knowing a patient's
age and using it as a distance between patients. If a patient's age is
important to predicting Alzheimer's disease the AIBD should outperform
the IBP; however, if the information is not informative no benefit should
be expected.\looseness=-1\vadjust{\goodbreak}

The comparison between the models with the AIBD and IBP priors uses the
same scheme as in the previous subsections. The prior for the mass parameter
in both models is a gamma$(1,1)$ distribution. Additionally for the AIBD
model, a gamma$(1,1)$ prior is set on the temperature parameter and a uniform
prior on the permutation. Again, we recommend the permutation parameter
be integrated out of the model, unless there is meaningful order of the
data. To tune the proposal distribution of the permutation parameter, we
set $k_{\rho }=8$ which produced an acceptance rate of approximately 20\%
in that Metropolis-Hastings step.

For posterior results we get, DIC values for the AIBD and the IBP models
are 32477.35 and 32539.26, respectively. Additionally, the average number
of customers per feature in the posterior samples is 74 for the AIBD and
70 for the IBP. The average number of customers per feature is 4.6 for
the AIBD and 4.1 for the IBP. The resulting posterior expected AUC for
each model is 0.759 and 0.747 for the AIBD and IBP, respectively. A 95\%
confidence interval on the paired difference between average AUC for the
AIBD and the IBP is $(0.011, 0.013)$. The results are fairly convincing:
the AIBD can use the extra distance information to obtain better classification
results.\looseness=1

This example demonstrates that a patient's age does contain information
that is valuable in classifying patients into healthy and Alzheimer's diseased
states. It also shows that the AIBD prior is able to capture this distance
information contained in a patient's age to improve model performance.

An informal speed comparison of the posterior sampling algorithms for the
IBP and the AIBD showed that the IBP could do one update of all parameters
in roughly 63 seconds, while the AIBD took approximately 70 seconds. This
comparison is easier to make than with the dd-IBP, since the algorithm
and software implementation are the same. This application demonstrates
that the computational penalty for using the nonexchangeable AIBD prior
isn't high, and improves model performance over the IBP prior.

\subsection{Items for Consideration}
\label{Sec:app-consider}

Procedures to check the convergence of the MCMC chains are rather standard
for most parameters of the model, however, since $\bm{\rho }$ is fairly
novel, we highlight a few considerations for it. As with essentially any
parameter, we cannot visit all possible values for $\bm{\rho }$ in posterior
sampling, and we treat it much the same as any other parameter. Although
we set a uniform prior on $\bm{\rho }$, we do not expect its posterior to
be uniform. What we want to accomplish is obtain a sample which provides
reasonable Monte Carlo estimates. We look for good acceptance rates and
exploration around the space by first checking that the chaining is moving
and, secondly, converging. To check for movement, we observe the standard
deviation of each item's permutation in each chain and ensure it is well
above 0. To check for convergence, we monitor the mean of the permutation
of the first half of the observations and we also monitor the mean of the
permutation of the odd observations. If these two summaries of
$\bm{\rho }$ show convergence, this provides evidence that the chain is
sampling from the stationary distribution. Additional summaries of
$\bm{\rho }$ may be warranted if concern about the mixing of
$\bm{\rho }$ exists.

The scalability of sampling from feature allocation models seems to be
somewhat limited from small to moderate $N$. Conducting a simple test,
we ran the AIBD model with a random temperature and permutation using three
different data sets with $N \in \{31,62,124\}$. The test first used
$\bm{X}_{31 \times 224}$, next $\bm{X}_{62 \times 224}$, and finally
$\bm{X}_{124 \times 224}$. The average time it took for one Gibbs scan
took on average 3.7, 20.4, and 195.5 seconds, respectively. All else being
equal, as $N$ grows so will the number of features $K$. The average
$K$ for those 3 tests are: 25.7, 36.7, and 78.8. We also ran the same tests
for the IBP and the average time it took for one Gibbs scan took on average
3.7, 20.9, and 188.7 seconds and the average $K$ for those 3 tests are:
25.7, 36.7, and 78.6, respectively. Therefore we see that the AIBD and
the IBP scale similarly. We suppose this occurs because the likelihood
computations dominate those of the prior. We expect the dd-IBP would scale
similarly for the same values of $N$. For larger $N$, the computation required
to get samples from the posterior might be too computationally expensive.
Thus other approaches might be required, e.g., variational inference. A
few works on variational inference in feature allocation models include
\cite{pmlr-v2-teh07a}, \cite{pmlr-v5-doshi09a}, and
\cite{pmlr-v33-ranganath14}.

The similarity function, which transforms distances into similarities,
is an important modeling choice in the AIBD. We considered a slightly altered
AIBD model, again with random temperature and permutation, but only changing
the similarity function. The similarity function we chose for the comparison
was the reciprocal function $1/(d_{i,j}+10^{-5})$. The results were fairly
similar to those obtained using the exponential decay similarity function.
This model has a DIC of 32539.38, a 95\% Monte Carlo (MC) confidence interval for AUC
is (0.751, 0.752) and for posterior expected number of features is (71.7,
71.8), and the average number of customers per feature in the posterior
samples is 4.5. In this case, the model does not seem to be overly sensitive
to the choice of the similarity function.\looseness=-1\vspace*{-3.5pt}

\section{Conclusion}
\label{discussion}\vspace*{-3.5pt}

In this paper, a generalization of the IBP was developed to allow for customer
dependence in the prior. It was demonstrated that after including pairwise
distance information, the AIBD preserves many familiar properties of the
IBP. We compared these properties of the AIBD to those of the IBP and dd-IBP,
and summarized them in Table~\ref{prior_summary_table}. Further, while the
AIBD and dd-IBP attempt to generalize the IBP by including pairwise distance
information, we have shown that the AIBD possesses several properties that
make it particularly appealing. An instance of this was shown in the application
in Section~\ref{dataAnalysis}, where the AIBD outperformed the dd-IBP in
terms of AUC. We note that the AIBD is constrained to cases where distances
are symmetric, whereas the dd-IBP allows for non-symmetric distances.

Overall, the AIBD is an attractive solution for incorporating distance
information into a prior distribution of a feature allocation. It retains
many desirable properties of the IBP. For a fixed mass and temperature,
it encourages more feature sharing between customers than the dd-IBP. Priors
can readily be set on model parameters, such as the temperature. Last,
but not least, this distribution and some associated methods are implemented
in the \textit{aibd} package in R \citep{AIBDpackage}.

\begin{acknowledgement}
The authors thank the editor, associate editor, and two anonymous referees
for their insightful comments that substantially improved this work.  This work was supported, in part, by NIH NIGMS R01 GM104972.
\end{acknowledgement}

\renewcommand{\thesection}{S.\arabic{section}}
\renewcommand{\thefigure}{S.\arabic{figure}}
\renewcommand{\thetable}{S.\arabic{table}}
\renewcommand{\theequation}{S.\arabic{equation}}
\setcounter{section}{0}
\setcounter{equation}{0}
\setcounter{table}{0}
\setcounter{figure}{0}

\def\thesection{S}
\newpage
\section*{Supplementary Material}

\subsection{The AIBD's Expected Number of Features per Observation. A~Proof that the AIBD has the Same Number of Expected Features as the IBP}
\label{sec:aibd-exp-features-per-row}

Here, we show that the sum of each row in the feature allocation matrix
from the AIBD (conditioned on the permutation $\bm{\rho }$, temperature
$\tau $, and mass $\alpha $) has an expected value of $\alpha $. Recall
that $X_{i}$ is the number of new dishes customer $i$ takes, and
$Y_{i}$ is the number of dishes sampled before customer $i$ samples dishes.
Let $Q_{i}$ be the number of previously-samples dishes that $i$ takes.
Let $S_{i} = X_{i} + Q_{i}$ be the total number of dishes sampled by customer
$i$. That is, $S_{i}$ is the sum of row $i$ in the feature allocation matrix
$Z$. We will show that
$E[S_{i} \mid \alpha ,\bm{\rho },\tau ]=\alpha $, and thus,
$E[S_{i} \mid \alpha ]=\alpha $, for all $i$. We will show this by strong
mathematical induction.

First, note that $S_{1} = X_{1}$, and
$X_{1} \mid \alpha , \bm{\rho }, \tau \sim \text{Poisson}(\alpha )$. Thus,
$E[S_{1}\mid \alpha , \bm{\rho }, \tau ] = E[X_{1}\mid \alpha ,
\bm{\rho }, \tau ] = \alpha $. Next, $S_{2} = X_{2} + Q_{2}$. Thus,
$E[S_{2} \mid \alpha , \bm{\rho }, \tau ] = E[X_{2} + Q_{2} \mid
\alpha , \bm{\rho }, \tau ] = E[X_{2} + X_{1,2} \mid \alpha , \bm{\rho },
\tau ] = E[X_{2} \mid \alpha , \bm{\rho }, \tau ] + E[X_{1,2} \mid
\alpha , \bm{\rho }, \tau ] = \alpha /2 + \alpha /2 = \alpha $, where
$X_{1,2}$ is the number of shared features between customers 1 and 2. (See \ref{aibd_enumf} for why
$E[X_{1,2} \mid \alpha , \bm{\rho }, \tau ] = \alpha /2$.)

Now, suppose that for some $i\in \mathbb{N}$,
$E[S_{j} \mid \alpha ,\bm{\rho }, \tau ] = \alpha $ for all $j \leq i$. Let
$C_{i} = \sum _{j=1}^{i} f(\tau , d_{\rho _{j}, \rho _{i+1}})$, then
\begin{eqnarray*}
E[S_{i+1} \mid \alpha , \bm{\rho }, \tau ] &=& E\left [X_{i+1} + \sum _{k=1}^{Y_{i+1}}
z_{i+1, k} ~\bigg |~ \alpha , \bm{\rho }, \tau \right ]
\\
&=& \frac{\alpha }{i + 1} + E\left [\sum _{k=1}^{Y_{i+1}} z_{i+1, k} ~
\bigg |~ \alpha , \bm{\rho }, \tau \right ]
\\
&=& \frac{\alpha }{i + 1} + E\left [\sum _{k=1}^{Y_{i+1}}
\frac{h_{i+1,
k}(\tau )\cdot i}{i + 1} ~\bigg |~ \alpha , \bm{\rho }, \tau \right ]
\\
&=& \frac{\alpha }{i + 1} + \frac{i}{i + 1} \cdot E\left [\sum _{k=1}^{Y_{i+1}}
h_{i+1, k}(\tau ) ~\bigg |~ \alpha , \bm{\rho }, \tau \right ]
\\
&=& \frac{\alpha }{i + 1} + \frac{i}{i + 1} \cdot E\left [\sum _{k=1}^{Y_{i+1}}
\frac{\sum _{j=1}^{i} f(\tau , d_{\rho _{j},
\rho _{i+1}}) \cdot z_{j,k}}{\sum _{j=1}^{i} f(\tau , d_{\rho _{j}, \rho _{i+1}})}
~\bigg |~ \alpha , \bm{\rho }, \tau \right ]
\\
&=& \frac{\alpha }{i + 1} + \frac{i}{i + 1} \cdot E\left [\sum _{k=1}^{Y_{i+1}}
\sum _{j=1}^{i} \frac{f(\tau , d_{\rho _{j},
\rho _{i+1}})}{C_{i}} \cdot z_{j,k} ~\bigg |~ \alpha , \bm{\rho }, \tau
\right ]
\\
&=& \frac{\alpha }{i + 1} + \frac{i}{i + 1} \cdot E\left [\sum _{j=1}^{i}
\frac{f(\tau , d_{\rho _{j},
\rho _{i+1}})}{C_{i}} \cdot \sum _{k=1}^{Y_{i+1}} z_{j,k} ~\bigg |~
\alpha , \bm{\rho }, \tau \right ]
\\
&=& \frac{\alpha }{i + 1} + \frac{i}{i + 1} \cdot \sum _{j=1}^{i}
\left ( \frac{f(\tau , d_{\rho _{j}, \rho _{i+1}})}{C_{i}} \cdot E
\left [ \cdot \sum _{k=1}^{Y_{i+1}} z_{j,k} ~\bigg |~ \alpha ,
\bm{\rho }, \tau \right ] \right )
\\
&=& \frac{\alpha }{i + 1} + \frac{i}{i + 1} \cdot \sum _{j=1}^{i}
\left ( \frac{f(\tau , d_{\rho _{j}, \rho _{i+1}})}{C_{i}} \cdot E[S_{j}
\mid \alpha , \bm{\rho }, \tau ] \right )
\\
&=& \frac{\alpha }{i + 1} + \frac{i}{i + 1} \cdot \sum _{j=1}^{i}
\left ( \frac{f(\tau , d_{\rho _{j}, \rho _{i+1}})}{C_{i}} \cdot
\alpha \right )
\\
&=& \frac{\alpha }{i + 1} + \frac{i}{i + 1} \cdot 1 \cdot \alpha
\\
&=& \alpha .
\end{eqnarray*}
Thus, $E[S_{i+1} \mid \alpha ,\bm{\rho }, \tau ] = \alpha $ also. Therefore,
by mathematical induction,
$E[S_{i} \mid \alpha ,\bm{\rho }, \tau ] = \alpha $, for all
$i \in \mathbb{N}$. Since for all $(\bm{\rho }, \tau )$,
$E[S_{i} \mid \alpha , \bm{\rho }, \tau ] = \alpha $, we also obtain
$E[S_{i} \mid \alpha ] = \alpha $.

\vspace*{-2pt}
\subsection{Simulation of MCMC Algorithm Accuracy. A~Posterior Simulation Study to Show the Accuracy of the Proposed MCMC Algorithm}
\label{MCMC_Simulation}\vspace*{-2pt}

We conducted a Monte Carlo study to demonstrate that the proposed MCMC
algorithm in Section~\ref{post_Z} approximately samples from the target
distribution. Recall that our algorithm has a truncation parameter, designed
to provide feasible computation with reasonable accuracy. We consider our
sampling algorithm for various truncation parameters and use our MCMC algorithm
to sample $\bm{Z}$'s from the AIBD prior distribution. We also sample
$\bm{Z}$'s using the constructive definition of the AIBD from Section~\ref{sec:aibd}. We then compare Monte Carlo estimates of univariate summaries
of the distribution and compare them to theoretical values. The default
value for the truncation parameter is $1000$ and we show that Monte Carlo
error dominates the small truncation error.

The experiment has two measures of accuracy. The first finds the largest
absolute difference of the empirical probability of getting a
$\bm{Z}$ with $K$ features (for all possible $K$) and its actual probability.
Recall $\bm{Z}$ is an $N\times K$ matrix, thus for $m$ samples we compute\looseness=-1
\begin{equation}
\label{eq:avgfeaterror}
\max _{\text{all }k} \ \Bigg | \, P(K=k) - \frac{1}{m} \sum _{i=1}^{m} I(K_{i}=k)
\, \Bigg |.
\end{equation}\looseness=0
The second measure of accuracy compares the average number of active features
with the theoretical results. This is expressed as
\begin{equation}
\label{eq:avgactivefeaterror}
\Bigg | \, \alpha N - \frac{1}{m} \sum _{i=1}^{m} tr(\bm{Z}_{i} {
\bm{Z}_{i}}^{T}) \, \Bigg |.
\end{equation}
It would be more accurate to compare the number of times each
$\bm{Z}$ was sampled with its theoretical probability, however, the space
of all possible $\bm{Z}$ is too large, even for moderate $N$, to take this
approach.

\begin{table*}[t]
\begin{tabular}{@{}|@{\ } l | c | c@{\ }|@{}}
\hline
Sampling Method & Max.\ Prob.\ of Feature Error & Mean Active Feature Error \\ \hline
Monte Carlo & 0.00024 & 0.0030 \\
MCMC (trunc.\ $=$ 1000) & 0.00024 & 0.0035 \\
MCMC (trunc.\ $=$ 100) & 0.00052 & 0.0102 \\
MCMC (trunc.\ $=$ 10) & 0.00210 & 0.1106 \\
MCMC (trunc.\ $=$ 1) & 0.00203 & 0.1255 \\ \hline
\end{tabular}
\caption{This table provides two metrics on the accuracy of the proposed MCMC
sampler from Section~\ref{post_Z}. It shows that the MCMC sampler with the
truncation parameter set to 1000 is nearly as accurate as a Monte Carlo sample
from the AIBD. As the truncation parameter decreases, so does the accuracy.}
\label{MCMCalgoGood}
\end{table*}

\begin{table}[t!]
\centering
{\footnotesize
\begin{tabular}{|r|c|cccccccccc|}
\hline
& $\tau $ & 1 / 2 & 1 / 3 & 1 / 4 & 1 / 5 & 2 / 3 & 2 / 4 & 2 / 5 & 3 / 4 & 3 / 5 & 4 / 5 \\ \hline
Distance & -- & 0.12 & 0.66 & 3.74 & 3.78 & 0.59 & 3.65 & 3.69 & 3.32 & 3.46 & 1.01 \\ \hline
Expected & \multirow{2}{*}{$0.2$} & 0.54 & 0.53 & 0.48 & 0.47 & 0.53 & 0.48 & 0.48 & 0.48 & 0.48 & 0.53 \\
Similarity & & 0.98 & 0.88 & 0.47 & 0.47 & 0.89 & 0.48 & 0.48 & 0.51 & 0.50 & 0.82 \\ \hline
Expected & \multirow{2}{*}{$1.0$} & 0.65 & 0.61 & 0.39 & 0.39 & 0.61 & 0.39 & 0.39 & 0.41 & 0.40 & 0.67 \\
Similarity & & 0.89 & 0.51 & 0.02 & 0.02 & 0.55 & 0.03 & 0.02 & 0.04 & 0.03 & 0.36 \\ \hline
Expected & \multirow{2}{*}{$5.0$} & 0.72 & 0.59 & 0.35 & 0.35 & 0.61 & 0.36 & 0.36 & 0.40 & 0.39 & 0.73 \\
Similarity & & 0.55 & 0.04 & 0.00 & 0.00 & 0.05 & 0.00 & 0.00 & 0.00 & 0.00 & 0.01 \\ \hline
\end{tabular}}
\caption{This table shows the expected number of features between two states
compared with the similarity of those two states, for three different
temperatures. As the temperature increases, the similarity (for this similarity
function) decreases and the sharing of features also changes, it goes up for
closer items and goes down for items that are relatively far apart.}
\label{Tab:ExpectVsSim}
\end{table}

\begin{figure}[t!]
\includegraphics[width=\textwidth]{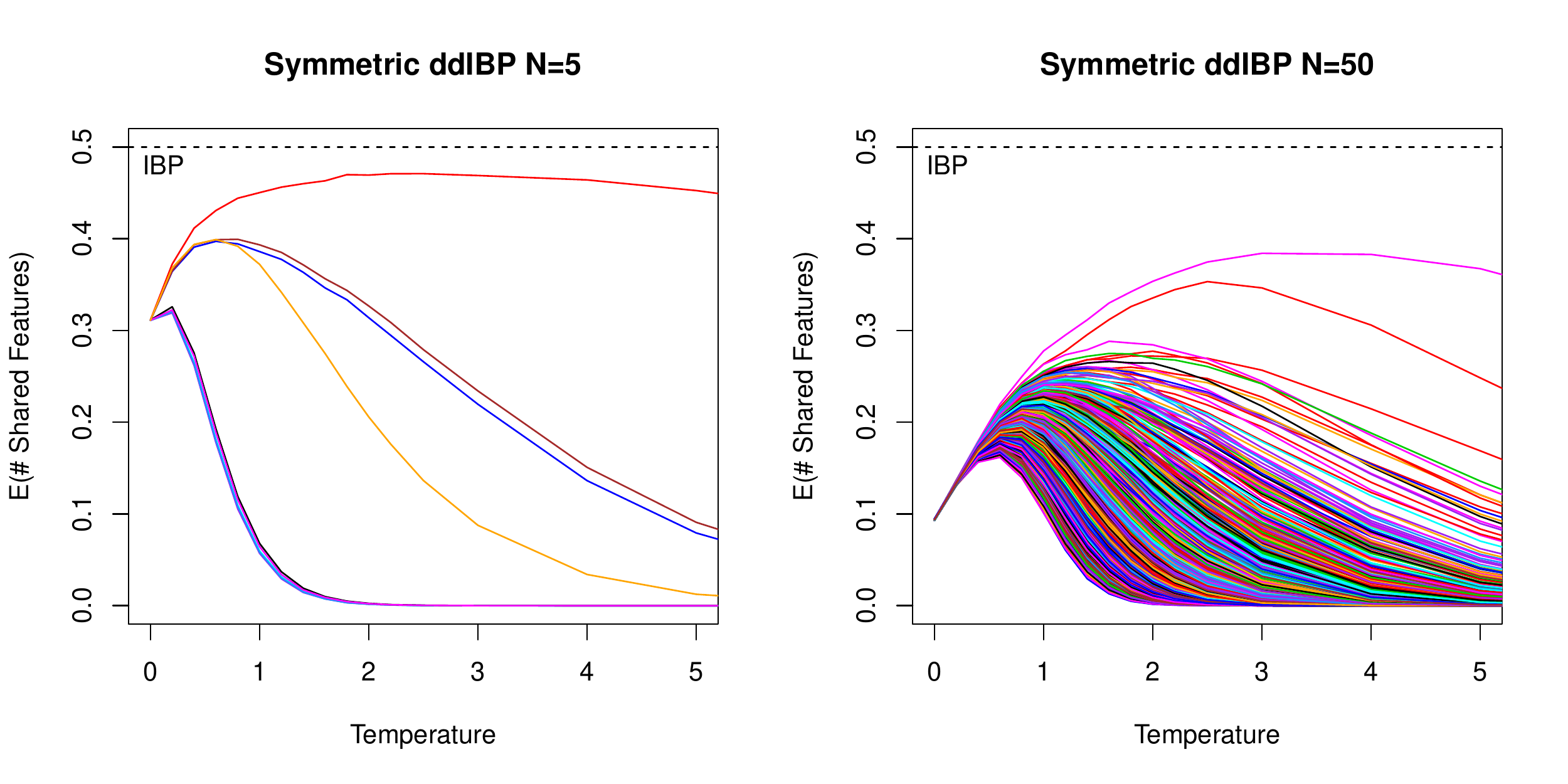}
\caption{The expected number of shared features between two states as a function
of temperature for the dd-IBP with a symmetric proximity matrix. In this case,
the permutation is irrelevant because there is no ordering of the customers. As
shown, the expected number of shared features has an initial increase and then
decreases as the temperature parameter increases. Across all pairs of states and
for all temperatures the amount of sharing is lower than the IBP.}
\label{symDDIBPFeats}\vspace*{-12pt}
\end{figure}

\begin{figure}[t!]
\includegraphics[width=\textwidth]{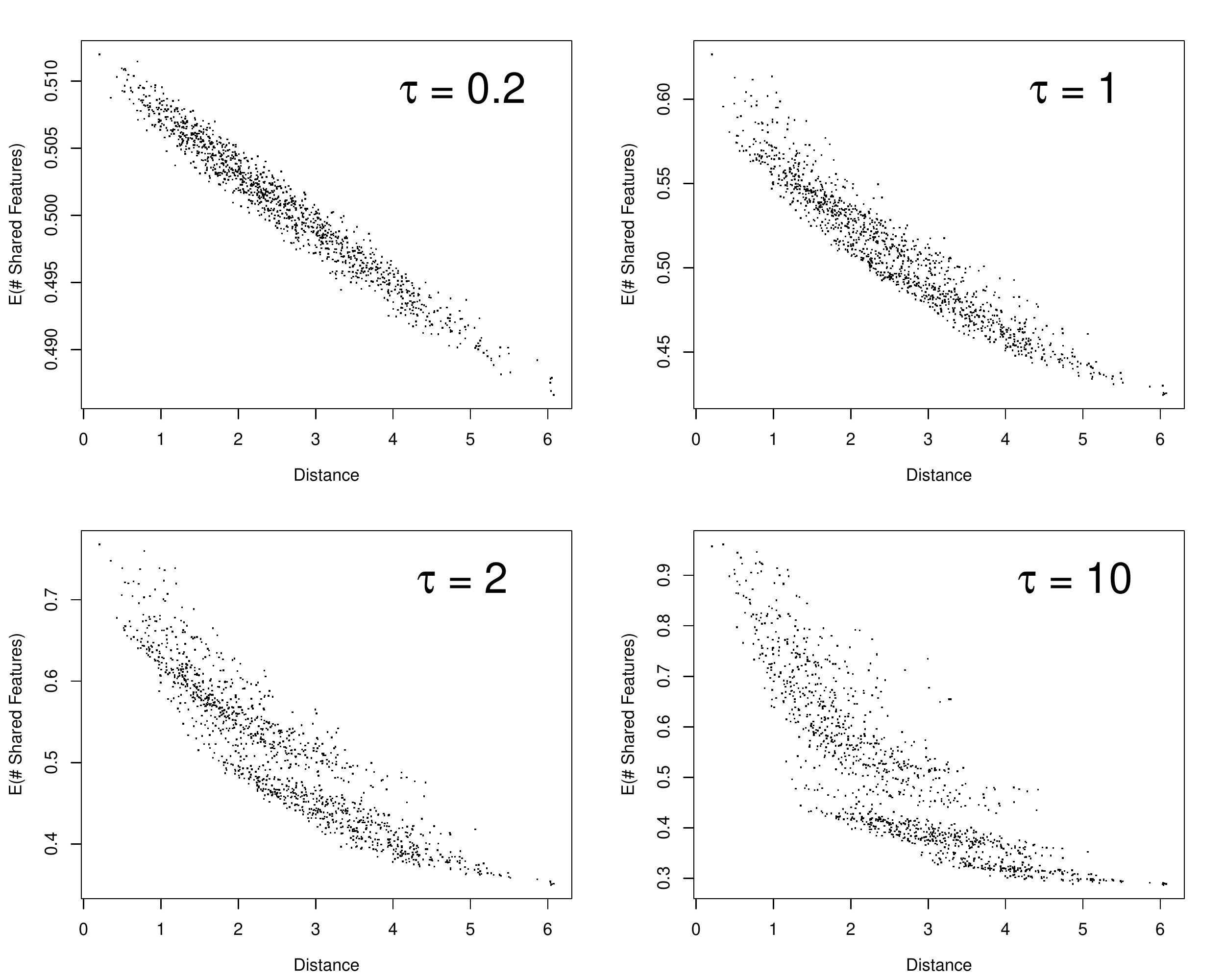}
\caption{These four plots show the pairwise distances against the expected
number of shared features in the AIBD when N$=$50 for different temperatures.
The permutation parameter is integrated out (using Monte Carlo).}
\label{AIBDScatPermInt}
\end{figure}

\begin{figure}[t!]
\includegraphics[width=\textwidth]{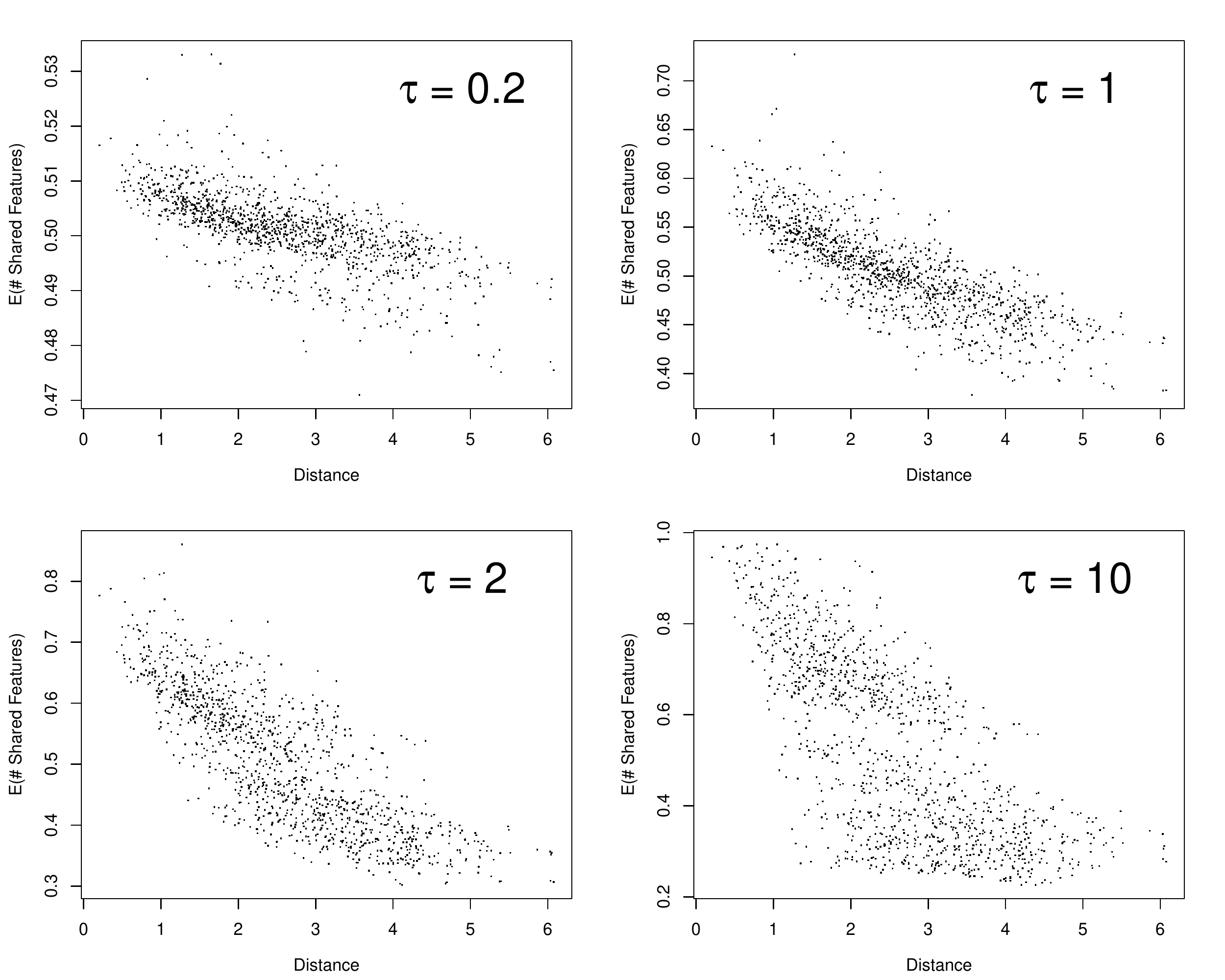}
\caption{These four plots show the pairwise distances against the expected
number of shared features in the AIBD when N$=$50 for different temperatures.
The permutation parameter is fixed using the alphabetical ordering of the
states.}
\label{AIBDScatPermFix}
\end{figure}

In this simulation we set $N=10$ and $\alpha =1.4$. The distance matrix
is defined by $d(i,j) = |i-j|/N$, and we use\vadjust{\goodbreak} the exponential decay function
to transform the distances to similarities with $\tau =2$. The permutation
is fixed at $1,2,\ldots ,10$, and for thinning we retain every tenth sample.
We obtain a total of one million samples for each case. We sample directly
from the AIBD for one case and use the MCMC sampler to obtain posterior
samples (with the likelihood proportional to 1) in the four other cases.
Each of the four cases of the MCMC sampler use different truncation parameter
settings $(1000,100,10,1)$.

The results are shown in Table~\ref{MCMCalgoGood} where lower numbers represent
lower error. The ``Max Prob of Feature Error'' column contain values from
(\ref{eq:avgfeaterror}) and the ``Avg Active Feature Error'' column
contain values from (\ref{eq:avgactivefeaterror}). It appears
the second metric is more sensitive to deviations from the truth. In Table~\ref{MCMCalgoGood}
we also included a Monte Carlo estimate (sampling directly
from the AIBD) to help quantify the amount of MC error one might expect.
Clearly, the Monte Carlo sample, directly from the target distribution,
is most accurate to the theoretical results. Our MCMC algorithm with the
truncation parameter set to 1000 is also very close to the truth. When
the truncation parameter is 100 the algorithm loses more accuracy and degrades
as it decreases.

This experiment supports our claim that the proposed algorithm is sampling\break
from the posterior directly, but it has some inherent truncation error.
This error can be reduced by increasing the
\texttt{newFeaturesTruncationDivisor} argument in the\break  \texttt{aibd::samplePosteriorLGLFM}
function. The default is set at 1000,
where we have observed that this truncation error is much smaller than
typical Monte Carlo error. A rough estimate from this simulation indicates
that the computational expense from increasing the truncation parameter
from 100 to 1000 is about 15\%.

\subsection{Additional Material on Feature Sharing}
\label{Append:FeatSharing}

This section further explores the sharing of features beyond that of the
manuscript. The data is from the USArrests dataset contained in R. When
$N=5$, the states selected are New Hampshire, Iowa, Wisconsin, California,
and Nevada (in that order). When $N=50$ all 50 states are used in the example.

Table~\ref{Tab:ExpectVsSim} shows the Euclidean distance (in the centered
and scaled data) between two states. It also includes the expected number
of features between two states compared with the similarity of those two
states, for three different temperatures. As shown in the table, as the
temperature increases the similarity, using the exponential decay function
to transform distance to similarity, decreases. Also the expected number
of shared of features also changes as a function of temperature. When the
temperature goes up the sharing between ``close'' states increases, but
for states that are relatively ``not close'' the sharing decreases.

\begin{figure}[t!]
\includegraphics[width=\textwidth]{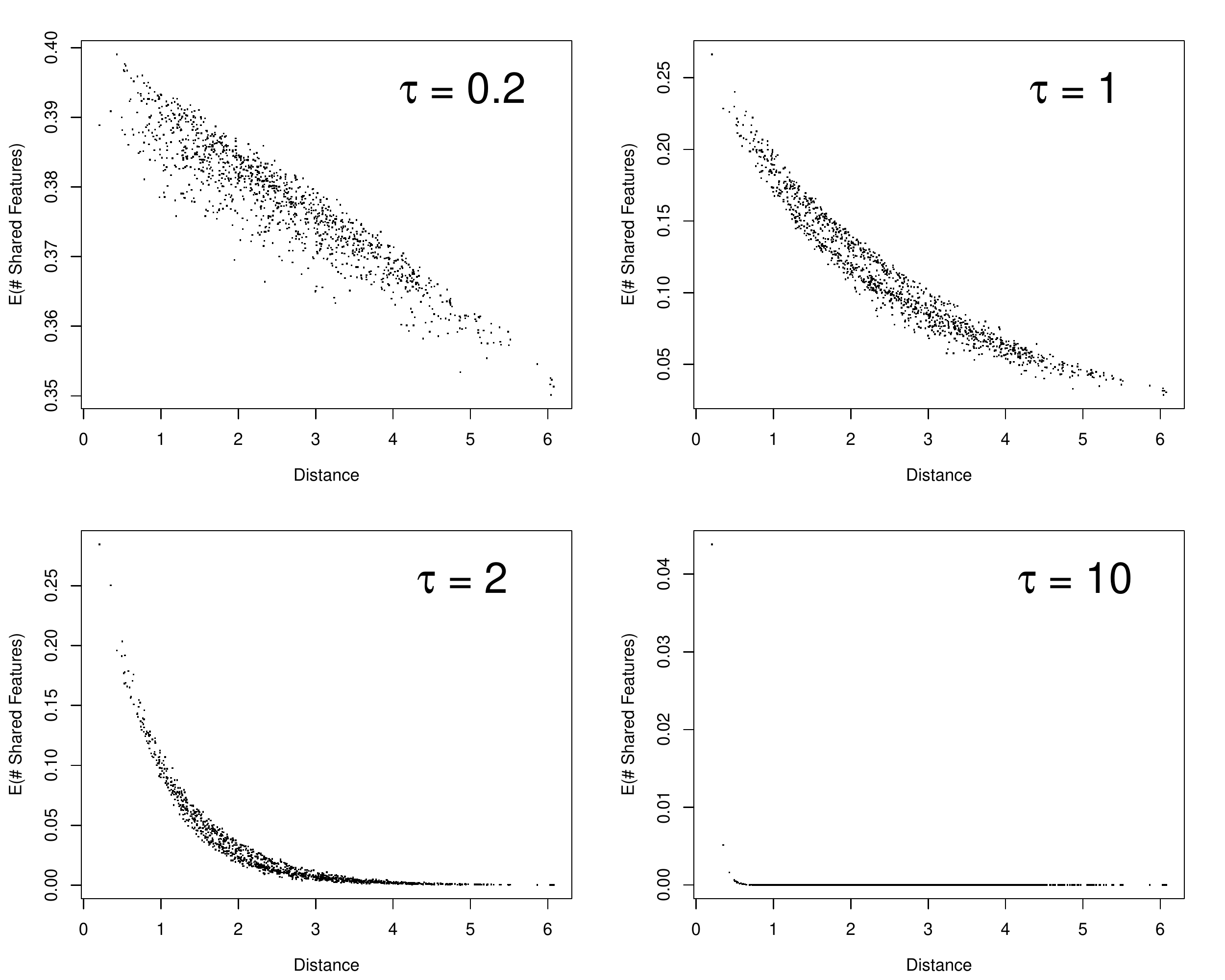}
\caption{These four plots show the pairwise distances against the expected
number of shared features in the dd-IBP when N$=$50 for different temperatures.
The permutation parameter is integrated out (using Monte Carlo).}
\label{ddIBPScatPermInt}
\end{figure}

\begin{figure}[t!]
\includegraphics[width=\textwidth]{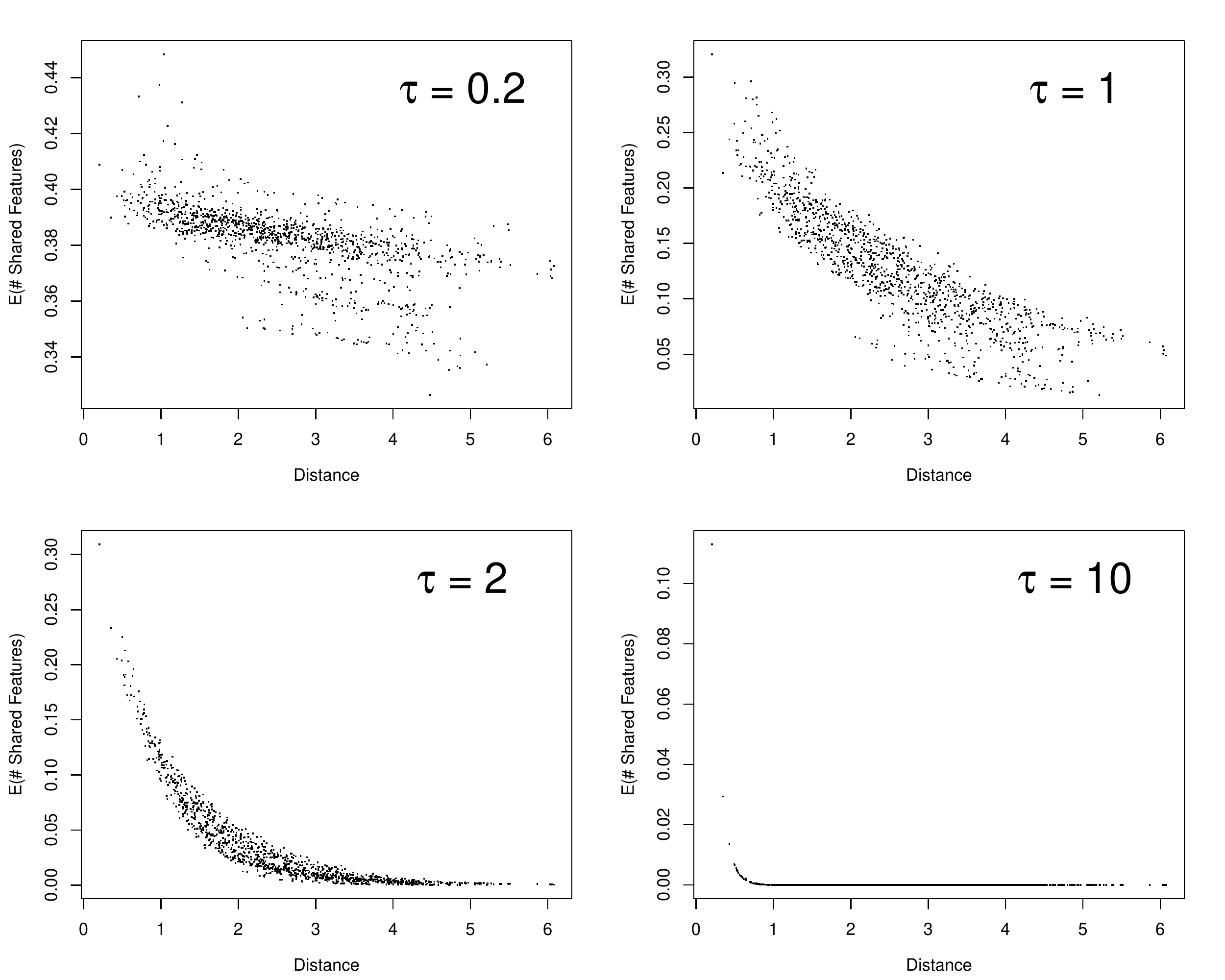}
\caption{These four plots show the pairwise distances against the expected
number of shared features in the dd-IBP when N$=$50 for different temperatures.
The permutation parameter is fixed using the alphabetical ordering of the
states.}
\label{ddIBPScatPermFix}\vspace*{-3.5pt}
\end{figure}

Figure~\ref{symDDIBPFeats} shows the expected number of shared features
between two states as a function of temperature for the dd-IBP with a symmetric
proximity matrix. In this case, the permutation is irrelevant because there
is no ordering of the customers. As shown, the expected number of shared
features has an initial increase and then decreases as the temperature
parameter increases. Across all pairs of states and for all temperatures
the amount of sharing is lower than the IBP (dashed line at 0.5). As the
temperature goes to infinity the sharing goes to zero.

Figures~\ref{AIBDScatPermInt}, \ref{AIBDScatPermFix}, \ref{ddIBPScatPermInt}, and \ref{ddIBPScatPermFix} explore the expected
pairwise feature sharing for the $50$ states in the USArrests dataset.
The plots show the expected shared number of features at four fixed temperatures,
where $\alpha =1$, for both the AIBD and the dd-IBP, and when the permutation
is averaged out of the model and fixed. As expected, there is a negative
correlation in all of the plots. When the temperature is zero all points
would be at 0.5 (with the IBP). At the temperature of 0.2 all four scenarios
show some change in feature sharing. As the temperature increases the AIBD
stays centered around 0.5, but the dd-IBP is driven towards no sharing.
Additionally, for both the AIBD and the dd-IBP the scatter of points becomes
less diffuse when the permutation is integrated out the model. In all cases
of the dd-IBP, all points are below the IBP level of 0.5.

\vspace*{-2pt}
\subsection{Posterior Summaries}
\label{Append:PostSums}\vspace*{-2pt}

This section contains the posterior means (and standard deviations in parentheses)
of the parameters for the various models from Section~\ref{dataAnalysis}. The first five rows of Table~\ref{Tab:PostSums} contain
the results for the AIBD with\vadjust{\goodbreak} fixed temperatures. The sixth row is from
the AIBD with random temperature and permutation. The seventh row of Table~\ref{Tab:PostSums} is the AIBD with random temperature and permutation
and using the reciprocal similarity function (all other models use the
exponential decay similarity function). Rows 8--12 show the results for
the dd-IBP models with fixed temperatures. The last row shows the posterior
summaries for the parameters in the IBP model.\vspace*{-3.5pt}

\begin{table*}[t]
\begin{tabular}{| l | c | c | c | c |}
\hline
& Mass & Temperature & $\sigma _{X}$ & $\sigma _{A}$ \\ \hline
AIBD (temp$=$0.4) & 12.1 (1.6) & 0.40 (0.00) & 0.52 (0.01) & 0.38 (0.01) \\
AIBD (temp$=$0.8) & 12.0 (1.6) & 0.80 (0.00) & 0.52 (0.01) & 0.37 (0.01) \\
AIBD (temp$=$1.2) & 11.8 (1.6) & 1.20 (0.00) & 0.52 (0.01) & 0.37 (0.01) \\
AIBD (temp$=$1.6) & 11.7 (1.6) & 1.60 (0.00) & 0.52 (0.01) & 0.37 (0.01) \\
AIBD (temp$=$2.0) & 11.6 (1.6) & 2.00 (0.00) & 0.52 (0.01) & 0.37 (0.01) \\
AIBD & 13.2 (1.7) & 2.77 (0.29) & 0.52 (0.01) & 0.35 (0.01) \\
AIBD (recip sim) & 12.7 (1.8) & 0.22 (0.02) & 0.52 (0.01) & 0.37 (0.01) \\
dd-IBP (temp$=$0.4) & 6.8 (1.7) & 0.40 (0.00) & 0.60 (0.03) & 0.32 (0.03) \\
dd-IBP (temp$=$0.8) & 5.8 (1.0) & 0.80 (0.00) & 0.58 (0.02) & 0.32 (0.02) \\
dd-IBP (temp$=$1.2) & 4.3 (0.9) & 1.20 (0.00) & 0.60 (0.02) & 0.35 (0.03) \\
dd-IBP (temp$=$1.6) & 5.0 (4.8) & 1.60 (0.00) & 0.59 (0.03) & 0.35 (0.02) \\
dd-IBP (temp$=$2.0) & 4.4 (2.6) & 2.00 (0.00) & 0.59 (0.03) & 0.36 (0.03) \\
IBP & 12.4 (1.6) & -- & 0.52 (0.01) & 0.38 (0.01) \\ \hline
\end{tabular}
\caption{Posterior parameter summaries. The posterior means of the parameters
for the various models. The posterior standard deviations are included in
parentheses.}
\label{Tab:PostSums}\vspace*{-3.5pt}
\end{table*}

\end{document}